\documentclass[fleqn,usenatbib]{mnras}

\usepackage{newtxtext,newtxmath}
\usepackage[T1]{fontenc}
\usepackage{ae,aecompl}

\usepackage{graphicx}	% Including figure files
\usepackage{amsmath}	% Advanced maths commands
\usepackage[utf8]{inputenc}
\usepackage{scrextend}
\usepackage{hyperref}
\title[ULXs in NGC 4861 and NGC 4449]{Optical Counterparts of ULXs in Two Dwarf Galaxies: NGC 4861 and NGC 4449}

\author[M. Ozdogan Ela et al.]{M. Ozdogan Ela,$^{1}$
A. Akyuz,$^{1,2}$\thanks{corresponding email:aakyuz@cu.edu.tr}
N. Aksaker,$^{1,3}$ S. Avdan,$^{1}$ İ. Akkaya Oralhan,$^{4}$ \and
A. Vinokurov,$^{5}$ S. Allak,$^{1,6}$ Y. Solovyeva,$^{5}$ K. Atapin,$^{7}$ and D. Bizyaev$^{5,7,8}$
\\ \\ 
$^1$Space Science and Solar Energy Research and Application Center (UZAYMER), University of Çukurova, 01330, Adana, Turkey\\
$^2$Department of Physics, University of Çukurova, 01330, Adana, Turkey\\
$^3$Adana Organised Industrial Zones Vocational School of Technical Science, University of Çukurova, 01410, Adana, Turkey\\
$^4$Department of Astronomy and Space Sciences, Erciyes University, 38039, Kayseri, Turkey\\
$^5$Special Astrophysical Observatory of the Russian AS, Nizhnij Arkhyz, Russia \\
$^6$Department of Physics, University of Çanakkale Onsekiz Mart, 17100, Çanakkale, Turkey\\
$^7$Sternberg Astronomical Institute,Moscow State University, Moscow, Russia\\
$^8$Apache Point Observatory and New Mexico State University, Sunspot, New Mexico, USA
}

\date{Accepted 2020. December 30; in original form 2020}
\pubyear{2020}

\begin{document}
\label{firstpage}
\pagerange{\pageref{firstpage}--\pageref{lastpage}}
\maketitle

% Abstract of the paper
\begin{abstract}
We present the results of a search for optical candidates of Ultraluminous X-ray sources (ULXs) in two dwarf galaxies: NGC 4861 and NGC 4449 using Hubble Space Telescope {\it HST} archival data. With a precise astrometry, we confirm that NGC 4861 X1 associated with an HII complex as reported by \cite{2014MNRAS.441.1841T} and we conclude that NGC 4861 X2 resides in a young star group with 400$\pm$80 M$\odot$. We also find that NGC 4449 X7 is associated with three optical candidates within an error radius of 0$\farcs$2 at the 90$\%$ confidence level. Absolute magnitudes (M$_{v}$) of these candidates are determined as $-$5.0 and $-$4.1. The age and mass values for the three candidates by using stellar evolutionary tracks are estimated as 40-50 Myr and $\sim$8 M$\odot$, respectively. The locations of optical candidates suggest possible association with a nearby group of stars. In addition, we analyzed the previously unused archival data of {\it XMM-Newton}, {\it Chandra} and {\it Swift} where the sources detected. Although the X-ray spectral data do not allow us to discriminate between physical models, long-term data at hand are consistent with the sources being in luminous hard states.
 
 %NGC 4861 X1 and NGC 4449 X1 show significant X-ray flux variability in factor of 5 and factor of 4, respectively. 

%even if the accretion disk contributes significantly to optical brightness

%within the error radius of 0$\farcs$28 and 0$\farcs$20 at the 90 $\%$ confidence level, respectively. 

%The spectral type for an optical candidate of NGC 4861 ULX-2 is B0I. The spectral types and ages of the optical candidates of NGC 4449 ULX-1 are between B0V $-$ B7V and 10 Myr $-$ 28 Myr, respectively.Color–magnitude diagrams have been used to investigate the properties of the candidates and their environments.%The use of a disk blackbody model for the mass of the compact objects indicates that these objects most likely have masses in the range 10 $-$ 20 M$\odot$.... .

\end{abstract}

% Select between one and six entries from the list of approved keywords.
% Don't make up new ones.

%%%%%%%%%%%%%%%%%%%%%%%%%%%%%%%%%%%%%%%%%%%%%%%%%%

%%%%%%%%%%%%%%%%% BODY OF PAPER %%%%%%%%%%%%%%%%%%

\section{Introduction}
\label{sec:intro}
Ultraluminous X-ray Sources are extragalactic non-nuclear sources with X-ray luminosities (L$_{X}$ > 10$^{39}$ erg s$^{-1}$) exceeding the Eddington limit for a stellar-mass black hole. Various models continue to be discussed in the literature about the nature of ULXs as a subclass of X-ray binary systems (see the review by \citealp{2017ARA&A..55..303K}). Studies of the last decade tend to favor that high luminosities of most ULXs can be explained by accretion at super-Eddington limits onto a central accretor which could be a stellar mass black hole or a neutron star \citep{2007MNRAS.377.1187P, 2015NatPh..11..551F, 2018ApJ...856..128W, 2018ApJ...857L...3W}. The high quality ULX spectra from {\it XMM-Newton} observations exhibit two component X-ray spectra with a soft excess and a turnover usually at 3-7 keV interval. These spectral features, unlike that were observed in Galactic black hole binaries (GBHBs), imply a new accretion regime and have been referred as the ultraluminous state that accompanies a super-Eddington accretion onto stellar-mass black holes \citep{2009MNRAS.397.1836G,2013MNRAS.435.1758S,2017ARA&A..55..303K}. On the other hand, detection of pulsations in 6 ULXs confirmed the presence of a neutron star as an accretor at least for a fraction of these systems such as 
ULX-2 in M82 \citep{2014Natur.514..202B}, ULX-1 in NGC 5907 \citep{2017Sci...355..817I}, P13 in NGC 7793 \citep{2016ApJ...831L..14F, 2017MNRAS.466L..48I}, ULX1 in NGC 300 \citep{2018MNRAS.476L..45C}, ULX-2 in NGC 1313 \citep{2019MNRAS.488L..35S} and ULX-7 in M51 \citep{2020ApJ...895...60R}. However, for some ULXs, sub-Eddington accretions onto intermediate mass black holes (IMBHs) are not completely ruled out for their high luminosity \citep{2004IJMPD..13....1M, 2009Natur.460...73F, 2015MNRAS.448.1893M}.

Unlike X-ray data, the number of ULXs defined by optical observations is still limited since these sources are quite faint in the optical (m$_{V}$ > 21 mag.) and are mostly located in crowded extragalactic regions. Determining the optical candidates and examining their properties to provide new constraints on the age, mass and possible spectral type of the donor star, make valuable contributions to ULX studies.
However, there are only few reliable optical counterparts known to date with spectral types determined by optical spectroscopy; for instance, a blue supergiant donor of type B9Ia for P13 in NGC 7793 \citep{2011AN....332..367M, 2014Natur.514..198M}, M-type supergiant donors for ULX-1 in NGC 253, ULX-2 in NGC 925, ULX-1 in NGC 4136 and ULX-1 in NGC300 \citep{ 2015MNRAS.453.3510H,2016MNRAS.459..771H, 2019ApJ...883L..34H}, Wolf-Rayet donors for ULX-1 in M101 \citep{2013Natur.503..500L} and for CG X-1 in Circinus \citep{2015MNRAS.452.1112E, 2019ApJ...877...57Q}.

The studies in the optical band showed that majority of the ULXs are found to be associated with star forming regions or star clusters \citep{2010MNRAS.403L..69P, 2011ApJ...734...23G, 2013MNRAS.432..506P, 2016ApJ...828..105A, 2019ApJ...875...68A, 2017MNRAS.467L...1E, 2018MNRAS.475.3561U}. These associations were supported by the counterparts of ULXs as early type OB stars due to their observed blue color \citep{2004ApJ...602..249L, 2005sf2a.conf..549G, 2005MNRAS.356...12S, 2008MNRAS.387...73R, 2013MNRAS.432..506P}. The blue color is thought to originate either from the X-ray radiation of the accretion disk or by a contamination from the accompanying star facing the X-ray source, or both \citep{2010MNRAS.403L..69P, 2012ApJ...745..123G, 2012ApJ...758...28J, 2018ApJ...854..176V}.
Furthermore, this association also implies that some of the donor stars could be red super giants, some of which are bright in the near-infrared H band indicating that the contribution of the accretion disc is lower in the near infrared than in the optical \citep{2016MNRAS.459..771H, 2019ApJ...883L..34H, 2017MNRAS.469..671L}.

In the present work, possible optical counterparts of three ULXs located in two dwarf galaxies: two of ULXs (X1 and X2) in NGC 4861 and one ULX (X7) in NGC 4449, have been searched extensively using the {\it HST} archival data. We also studied the X-ray spectral and temporal properties of these three ULXs. For these, we used 2018 {\it Chandra} and {\it XMM-Newton} data sets, respectively. These data were unused before. We also reanalysed the previous {\it XMM-Newton} and {\it Chandra} and {\it Swift} data.

NGC 4861 (Mrk 59) is classified as a Magellanic-type irregular galaxy at a distance of 9.95 Mpc \citep{2013AJ....146...86T} and is called a blue compact dwarf galaxy because of its blue color and high UV continuum emission \citep{1980ApJ...240...41F, 1981ApJ...247..823T, 1998ApJ...497..601K, 1999ApJ...511..639I, 2006A&A...459...71I, 2009ApJ...700..654E}. NGC 4861 was also classified as the first example of a "comet-like" galaxy by \cite{1986ApJ...309...59L}. The "comet's head" at the one end is characterized by a high surface brightness star-forming region and elongated low surface brightness area, the "tail", suggesting an irregular dwarf galaxy seen edge on.
Two X-ray sources IXO 72 and IXO 73 in NGC 4861 had already been identified as ULXs using {\it ROSAT} observations by \cite{2002ApJS..143...25C}. Also, based on {\it ROSAT} data, \cite{2005ApJS..157...59L} reported the sources IXO 73 and IXO 72 as X1 and X2 respectively, in their ULX catalog. The X-ray emission of these two sources were further investigated by \cite{2014MNRAS.441.1841T}. In their study, X1 and X2 were both associated with HII regions and bright massive stars with luminosities in 0.5$-$10 keV energy band as $\sim$(2.4-8.9)$\times$10$^{39}$ erg s$^{-1}$ for X1 and $\sim$(1.8–2.1)$\times$10$^{39}$ erg s$^{-1}$for X2, from the analysis of 2003 {\it XMM-Newton} and 2012 {\it Chandra} data.

NGC 4449 is also a Magellanic-type, irregular, star-forming galaxy at a distance of 4.02 Mpc \citep{2018ApJS..235...23S}. Using the {\it ROSAT} data, 7 point sources were detected within the 25th magnitude ellipse (D$_{25}$) of NGC 4449. The source X7 was identified as a ULX \citep{1997A&A...319..459V}. \cite{2005ApJ...633.1052F} determined its X-ray luminosity as 1.9$\times$10$^{39}$ erg s$^{-1}$ using {\it XMM-Newton} data (ObsID 0112521701). \cite{2006ApJ...649..730W} reported 3 bright X-ray point sources in NGC 4449 using the same {\it XMM-Newton} data. X7 was numbered as XMM1 in their list while the remaining two sources were identified as possible supernova remnants. The source X7 was also identified as a ULX with an X-ray luminosity of 1.5$\times$10$^{39}$ erg s$^{-1}$ using {\it Chandra} data by \cite{2011ApJ...741...49S}.

With this background, the present paper is organized as follows: X-ray and optical data reductions and details of data analysis are described in Section \ref{sec:obs}. A summary of the optical and X-ray properties of the optical candidates of ULXs and the interpretation of results and discussion are given in Section \ref{sec:res}.

\section{Data Reductions and Results}
\label{sec:obs}

The X-ray and optical properties of NGC 4861 X1, X2 and NGC 4449 X7 are examined in the light of data summarized Table \ref{T:tab1} and Table \ref{T:tab2}. 

\subsection{Observations, Data reductions and Analyses}
\subsubsection{Astrometry}
\textbf{NGC 4861}

\begin{figure*}
	\begin{center}
		\includegraphics[width=\textwidth]{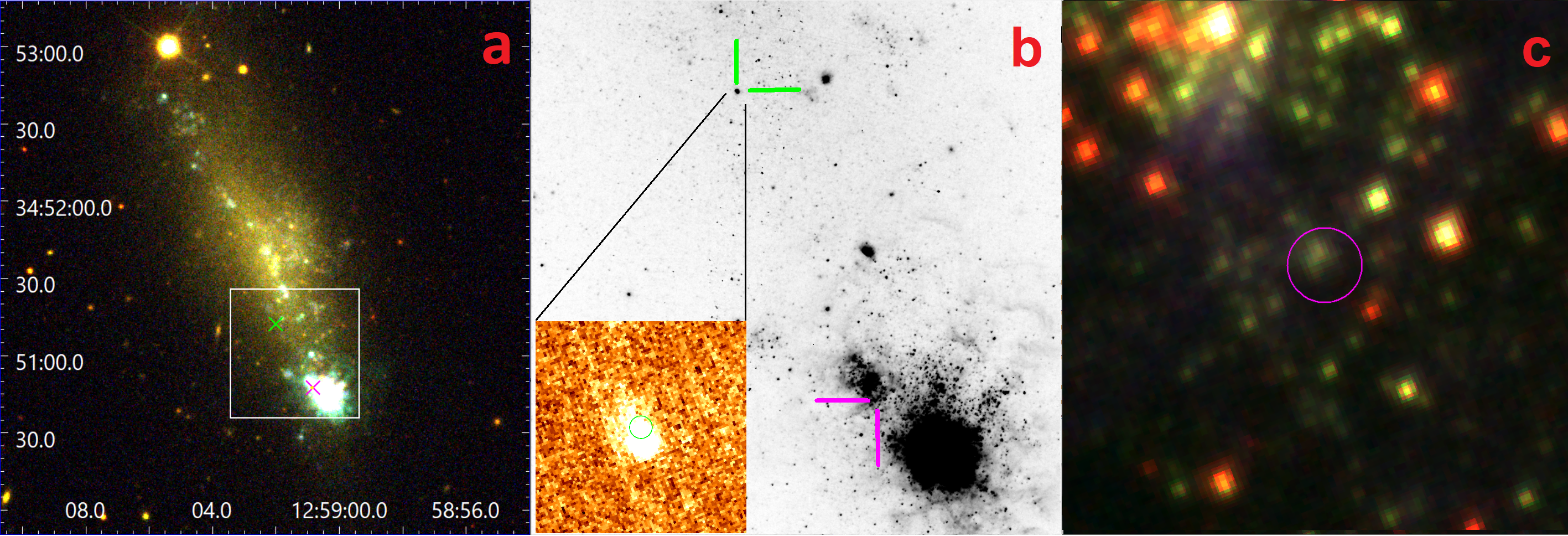}
		\caption{a) Three color {\it SDSS} image of NGC 4861 (red, green, and blue colors represent i, r and u bands, respectively) and the locations of X1 (green cross) and X2 (magenta cross) are marked within the white box (45$\arcsec$ $\times$45$\arcsec$). b) The green and magenta color lines show the position of X1 and X2 in {\it HST}/ACS/F606W, respectively. The green color circle shows the position of corrected X1 on the {\it HST}/WFC3/F658N image as inset. c) The magenta circle represents the error radius (0$^{\prime\prime}$.24) of X2 and possible optical counterpart(s) are shown on three color {\it HST} image of NGC 4861 (red, green, and blue colors represent F814W, F547M and F487N bands, respectively). The north is up in all images. X-axis and y-axis represent R.A. and Dec., respectively.}
		\label{Fig1}
	\end{center}
\end{figure*}

Identification of optical counterpart(s) of ULXs requires rather precise astrometry. For this, we matched the objects between {\it Chandra} and {\it HST} observations to identify possible optical counterparts of X1 and X2 in NGC 4861. For {\it Chandra}, the longest ACIS observation (ObsID 20992) was chosen and wavedetect task in the {\scshape ciao} was used for the X-ray source detection. Also, the task daofind in {\scshape iraf} was run for the optical source detection on ObsID jc9v72010 image of {\it HST}. We could not find any matching source between {\it Chandra} and {\it HST} images. Therefore, {\it GAIA} source catalog \citep{2018A&A...616A...1G} was used for astrometric correction with the {\it GAIA}-Graphical Astronomy and Image Analysis Tool \footnote{http://star-www.dur.ac.uk/$\sim$pdraper/gaia/gaia.html}. From this catalog, we selected stars that appear to be isolated and point like sources in {\it Chandra} and {\it HST} images. We found 3 and 5 reference sources in {\it Chandra}$-${\it GAIA} and {\it GAIA}$-${\it HST} images, respectively. The astrometric offsets between {\it Chandra} and {\it GAIA} were found as 1$\farcs$69 $\pm$ 0.08 for R.A. and 0$\farcs$20 $\pm$0.07 for Dec, while between {\it GAIA} and {\it HST} were found as 0$\farcs$15 $\pm$0.000022 for R.A. and 0$\farcs$04 $\pm$ 0.01 for Dec. with 1$\sigma$ errors. After astrometric correction, the positional uncertainties for X1 and X2 were obtained as 0$\farcs$24 at 90\% confidence level in ACS/F606W image. The corrected coordinates of ULXs were given in Table \ref{T:tab3}. The locations of two ULXs on the three-color {\it SDSS} image of NGC 4861 are shown in Fig. \ref{Fig1}a and positions of X1 and X2 are given on {\it HST}/ACS F606W filter Fig. \ref{Fig1}b.\\
\textbf{NGC 4449}

Relative astrometry between {\it Chandra}/ACIS-S (ObsID 10875) and {\it HST} (ObsID j9dr01010) observations were performed to identify optical counterpart(s) of X7 in NGC 4449. Using ACS/WFC/F550M image taken on November 18, 2005, we have found optical counterparts for five X-ray sources, their total photon counts are in the range of 20-1786 counts. The brightest of them is a well studied, young and very luminous oxygen-rich supernova remnant named as SNR4449-1 \citep{1998AAS...193.7404B, 2013MNRAS.436.2454M}, the nature of the remaining sources is not known. Both X7 and five reference sources are located on chip S3 with a relatively small offset ($<2^\prime$) from the optical axis of the instrument.
 
The X7 position derived from ACIS-S is R.A.= 12:28:17.831, Dec.= +44:06:34.10. These coordinates along with those of the reference sources are given in Table \ref{T:tab3b}. We accepted 1$\sigma$ accuracy in {\it Chandra} position as 0$\farcs$1. All five reference sources are clearly seen on ACS/WFC/F550M image, statistical errors for their positions are negligible ($\lesssim$0$\farcs$01).  Considering these sources, we found offsets between  {\it HST} and {\it Chandra} images as 0$\farcs$76 $\pm$ 0.05 in R.A., and 1$\farcs$05 $\pm$ 0.05 in Dec. After astrometric correction, the position of X7 on F550M image becomes R.A.= 12:28:17.901, Dec.= +44:06:35.15 with uncertainties of 0$\farcs$2 at 90\% confidence level (errors were combined in quadrature).

\begin{figure*}
	\begin{center}
		\includegraphics[width=\textwidth]{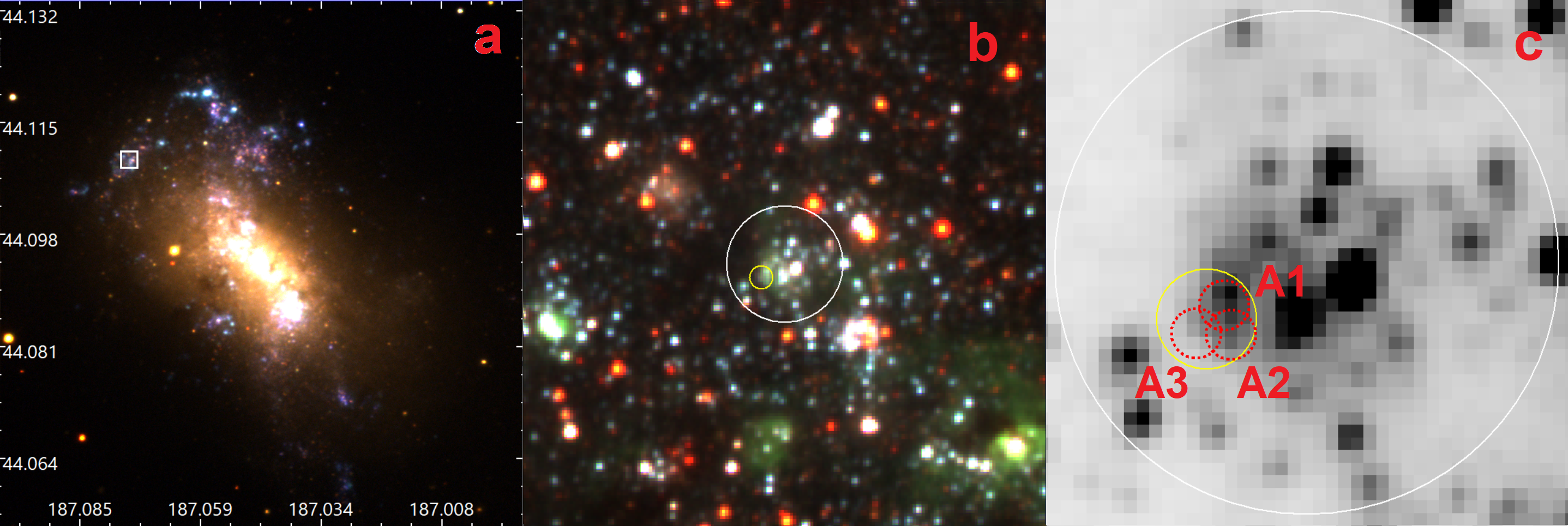}
		\caption{a) Three color {\it SDSS} image of NGC 4449 (red, green, and blue colors represent i, r and u bands, respectively). The white box (5$\arcsec$ $\times$5$\arcsec$) shows the location of ULX X7. b) Three color of {\it HST}/ACS/WFC image is zoomed onto the white box given in (a). The red, green, and blue colors represent F814W, F555W and F435W bands, respectively. 1$\arcsec$ radius (white circle) was chosen around the X7 and include star cluster investigated in CMD and the yellow circle represent error circle of 0$\farcs$2. c) Three optical candidates are labelled A1, A2 and A3 on {\it HST}/ACS F555W image.}
		\label{Fig2}
	\end{center}
\end{figure*}

The three color {\it SDSS} image of NGC 4449 with the (approximate) position of X7 is shown in Fig. \ref{Fig2}a. As a result of the astrometric correction, the location of X7 with surrounding region is shown in Fig. \ref{Fig2}b and more detailed image of this region with optical candidates can be seen in Fig. \ref{Fig2}c.

\subsubsection{Photometry}

The Point Spread Function (PSF) photometry was performed by using {\scshape dolphot} v2.0 package with ACS and WFC3 modules. The data were retrieved from the {\it HST} data archive\footnote{https://archive.stsci.edu/hst/search.php} (see Table \ref{T:tab1}), and reduction steps were followed as given in the {\scshape dolphot} manual \citep{2000PASP..112.1383D}. The images were processed by masking all bad pixels using the {\scshape acsmask} task and the files were split into single chip images using the {\scshape splitgroups} task before performing photometry. The sky background for each chip was calculated with the {\scshape calcsky} task. Then {\scshape dolphot} task was run for the detection of sources and their photometry. In order to eliminate fake sources around the position of ULXs, the value of {\it sigfind}=3.5 was used. 

After the photometry, we investigated optical counterpart(s) of NGC 4861 X1, X2 and NGC 4449 X7. The NGC 4861 X1 is associated with a bright extended region within 0$\farcs$24 error circle as seen in Fig. \ref{Fig1}b, which makes it very difficult to identify as a point-like counterpart. Therefore, we have not done any further optical analysis for NGC 4861 X1. On the other hand, a single but extended optical counterpart as a possible star cluster is identified within the error radius of 0$\farcs$24 for NGC 4861 X2 (see Fig. \ref{Fig1}c). The magnitudes in VegaMag for the candidate star cluster are given in Table \ref{T:tab4}. Both NGC 4861 X1 and X2 are coincident with a luminous HII region in the F658N image as noted in \cite{2014MNRAS.441.1841T}. In case of NGC 4449 X7, three candidates for optical counterparts are identified in the error circle with a radius of 0$\farcs$2 (see Fig. \ref{Fig2}c). Photometric analysis results of these optical candidates are given in Table \ref{T:tab5}.

\subsubsection{Spectral Observations}
The ground-based spectral observations of NGC 4449 and NGC 4861 were made with the SCORPIO instrument at the 6 m BTA telescope of the SAO RAS \citep{2005AstL...31..194A}. We obtained the spectra of the nebulae near the ULXs in both galaxies on March 22, 2020. Data reduction was carried out with the LONG context in {\scshape midas} using standard algorithm. The spectra were extracted and converted into 1D spectra using the {\scshape spextra} package \citep{2017AstBu..72..486S}. Wavelength calibrations were made using neon lamps. Spectral data of the standard stars \footnote{https://www.eso.org/sci/observing/tools/standards/spectra/okestandards\_rev.html} taken on the same night were used for the flux calibration.

For reddening measurements, we used H$_{\delta}$/H$_{\beta}$ and H$_{\gamma}$/H$_{\beta}$ line ratios. The extinction (A$_v$) toward the NGC 4861 X1 and X2 were found 0.35 ($+$0.14,$-$0.13) mag and 0.12 ($+$0.13,$-$0.10) mag, respectively. A$_v$ was also calculated as 0.56$\pm$0.14 mag for NGC 4449 X7. These values were used throughout this study.
 
\subsubsection {Age and Mass Estimation of Optical Counterpart of NGC 4861 X2}

The optical counterpart of X2 is seen as an extended source rather than a point-like one. We investigated whether this source could be a star cluster by determining its age and mass values. We used Starburst99 v7.0.1 software\footnote{https://www.stsci.edu/science/starburst99/docs/default.htm} \citep{1999ApJS..123....3L,2019A&A...629A.134G} to determine the age.
This software produces model spectra for continuum emission of star clusters as a function of the cluster age and metallicity for chosen evolution tracks. A total mass for these tracks was considered only as a normalization constant and was fixed at 10$^{6}$ M$\odot$. The Padova tracks with asymptotic giant branch stars were utilised for the metallicity of Z=0.0024 \citep{2014A&A...564A.121C}. The initial mass function (IMF) was identified by the Kroupa law with indices of 1.3 for stars of 0.1$-$0.5 M$\odot$ and 2.3 for 0.5$-$300 M$\odot$ \citep{2001MNRAS.322..231K}. We computed a grid of models for cluster ages ranging from 1 Myr to 1 Gyr with a step 5 Myr. Then convolving each model with the filter throughput (except F658N filter, see Table \ref{T:tab4}) and adjusting the normalization, we found that the best fit was achieved by the model of an age $\sim$ 5 Myr with a $\chi^{2}$ value of 5.28 for seven filters and two free model parameters.

Minimum mass of this extended object was obtained from the extinction-corrected V-band luminosity and the age dependent mass-to-light ratio predicted by the \cite{2003MNRAS.344..1000B} simple stellar population (SSP) models assuming a distance modulus of 29.62 mag and the metallicity of this galaxy. We selected two-part power-law IMF from SSP models and found a total mass as 400$\pm$80 M$\odot$ for this object. \cite{2013MNRAS.432..506P} found that all clusters associated with the ULXs are extremely young with the age of less than 6 Myr and their mass range are between 5$\times10^{3}$ M$_{\odot}$ and 6$\times10^{5}$ M$_{\odot}$ in Antennae galaxies. The age of 5 Myr we found is compatible with the young cluster hosting a ULX. However, the corresponding mass of 400$\pm$80 M$\odot$ is too small for such young clusters \citep{2011ApJ...741...86R, 2013MNRAS.432..506P}. Therefore, the optical counterpart of X2 could be a small group of stars rather than a huge star cluster.

\subsubsection {Color-Magnitude Diagram (CMD)}
Color-Magnitude Diagram (CMD) was obtained for the age estimation of optical candidates of NGC 4449 X7 assuming that the donor star dominates optical emission. For this, Padova Stellar Evolution Code (PARSEC; \citealp{2012MNRAS.427..127B}) isochrones were downloaded from the web\footnote{http://stev.oapd.inaf.it/cgi-bin/cmd}. These isochrones were derived based on the {\it HST}/ACS/WFC wide filters photometric system. 

\begin{figure*}
\begin{center}
		\includegraphics[width=\columnwidth]{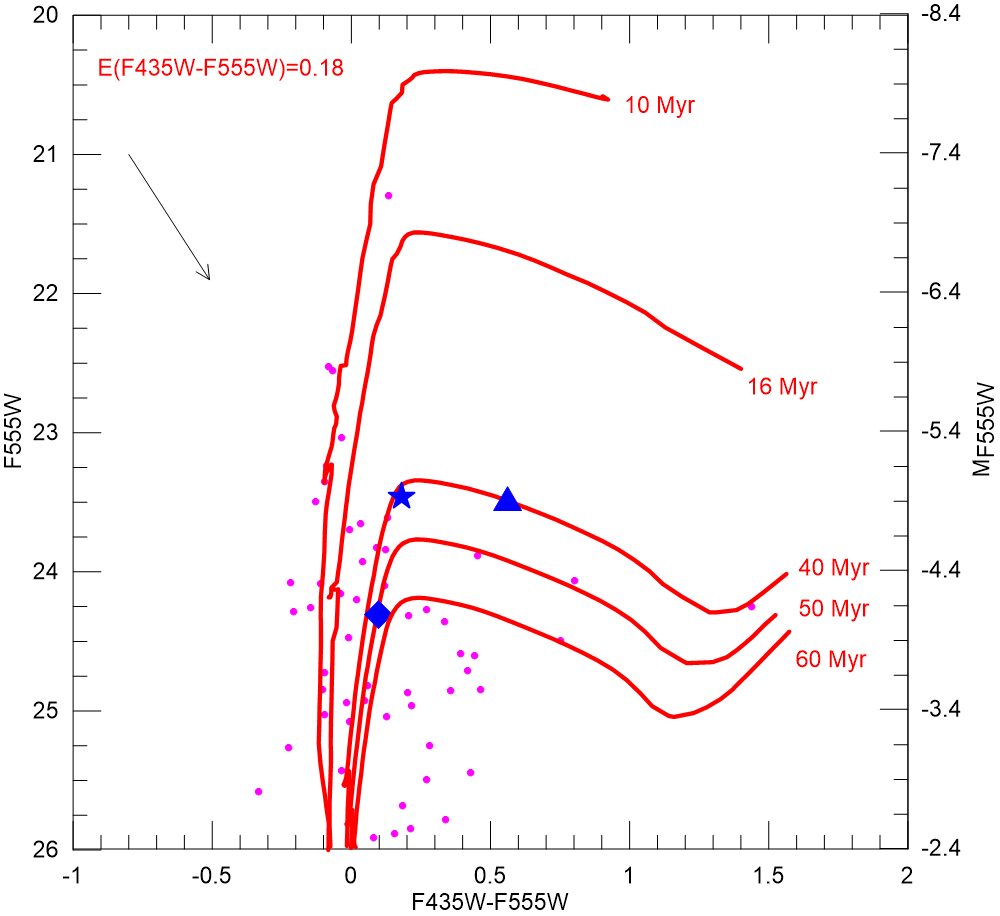}
		\caption{The {\it HST}/ACS color$-$magnitude diagram (CMD) for optical candidates of NGC 4449 X7 and star cluster around X7. Padova isochrones of different ages are overplotted. Triangle, star and diamond represent optical candidates A1, A2 and A3, respectively. The magenta dots show the members of star cluster selected by 1$\arcsec$ circle around the X7 shown in Fig. \ref{Fig2}b. These isochrones have been corrected for the extinction value of $A_{V}=0.56$ mag and black arrows show the reddening line.}
		\label{F:Fig3}
	\end{center}
\end{figure*}

\cite{1998ApJ...507..179H} derived internal reddening E(B$-$V) = 0.18 using H$_{\alpha}$/H$_{\beta}$ ratio measured from HII regions in NGC 4449. Also, the foreground reddening was given as 0.016 by \cite{2011ApJ...737..103S}. E(B$-$V) = 0.18 was taken as the average internal reddening for NGC 4449 since foreground reddening is low. This averaged value is in very good agreement with our calculated local value A$_v$, obtained for X7. The distance modulus (28.02 mag) and metallicity (Z=0.004; \citealp{2008AJ....135.1900A}) values have been used to produce the CMD. The resultant CMD is given in Fig. \ref{F:Fig3} by using {\it HST}/ACS F435W and F555W filters.

\subsection{X-Ray Observations}
The galaxies of NGC 4449 and NGC 4861 were observed by {\it Chandra} and {\it XMM-Newton} between 2001 and 2018. Also, {\it Swift} observations were obtained between 2007-2017 for NGC 4449. The log of X-ray observations is given in Table \ref{T:tab2}. The C2, C3, C4 and C5 observations, that were obtained in 10 days, for NGC 4861 and only XM6 for NGC 4449 have not been used for 3 ULXs in any previous analyses. Below we describe the data reduction steps applied to the analysis, for these sources in two galaxies.

{\it XMM-Newton} data reductions were carried out using the {\scshape sas} (Science Analysis Software version 18.0). The {\scshape epchain} and {\scshape emchain} tasks were used to obtain EPIC-pn and MOS event files for each observation. The events corresponding to PATTERN$\leq$12 and PATTERN$\leq$4 with FLAG=0 were selected for EPIC-MOS and pn cameras, respectively. Source and background spectra were extracted using the {\scshape evselect} task, with appropriate circular regions of radii $15\arcsec$ and $20\arcsec$, respectively. The background regions were, then, selected from source free regions on the same chip containing the ULXs. EPIC-pn and MOS spectra were fitted simultaneously in the 0.3$-$10 keV energy band. A constant scaling factor was taken in order to consider the cross calibration differences between the instruments. The files for the source and background spectra were grouped with a minimum of 20 counts per bin.

{\it Chandra} data reductions were performed using the {\scshape ciao} ({\it Chandra} Interactive Analysis of Observations, version 4.12) with the {\scshape caldb} (version 4.9.1). The source was located on the chip ACIS$-$S3. Data were reprocessed using {\scshape chandra\_repro} to produce up-to date event lists. The source and background photons were extracted with a {\scshape specextract} task using radius $3\arcsec$ circular regions. The spectra were grouped from 5 to 15 counts per bin.

Swift-XRT data sets were processed with the {\it xrtpipeline} along with HEASARC. Photon Counting (PC) data were used in the 0.3–10 keV energy range. The source and background photons were extracted circular regions of radii 15 and 30 arcsec, respectively. Appropriate ancillary response files were generated with the task {\scshape xrtmkarf}. We combined spectra from all 2013 observations for fitting purpose due to the short exposure time and low statistics. The spectra were grouped at 10 counts per bin.

{\scshape xspec} software (version 12.11.0) included in the {\scshape heasoft} (v6.27.2) package was used for spectral analysis of the X-ray data. The following models were applied to the spectra of 3 ULX sources: (1) Power-law ({\scshape pl}), (2) diskblackbody ({\scshape dıskbb}), (3) p- free diskblackbody model ({\scshape dıskpbb}), (4) {\scshape pl+dıskbb} and (5) {\scshape pl+dıskpbb}. Results are given in Table \ref{T:tab7} and Table \ref{T:tab8} with well-fitted models and their parameters. The spectral fits were performed using the C-statistics \citep{1979ApJ...228..939C} for goodness of fit tests due to the low number of photons available. As given in Table \ref{T:tab7} and \ref{T:tab8}, two component models do not improve the fit significantly compared to single component models. Moreover, for most of the datasets, the parameters of the two-component model parameters are not physically meaningful although they provide statistically acceptable fits to some of the data.

\begin{figure*}
\begin{center}
\includegraphics[width=\columnwidth]{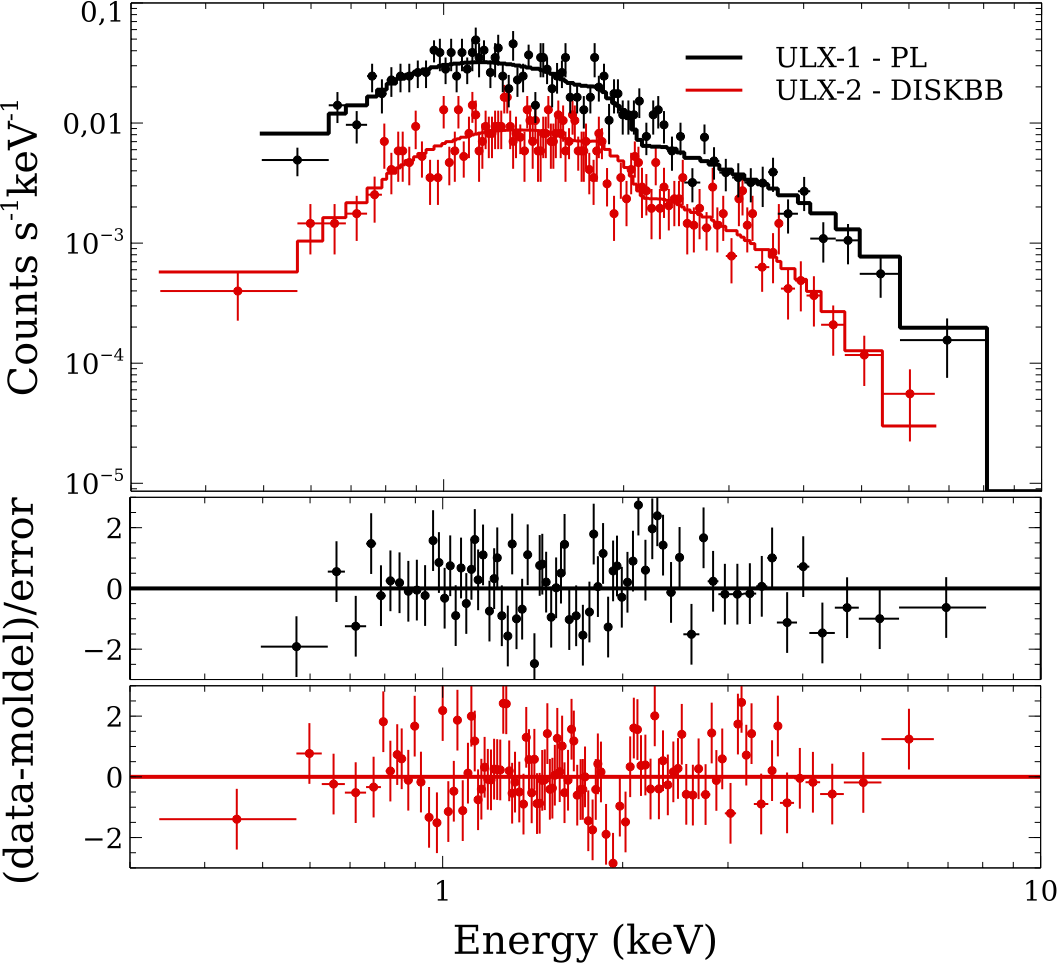}
\includegraphics[width=\columnwidth]{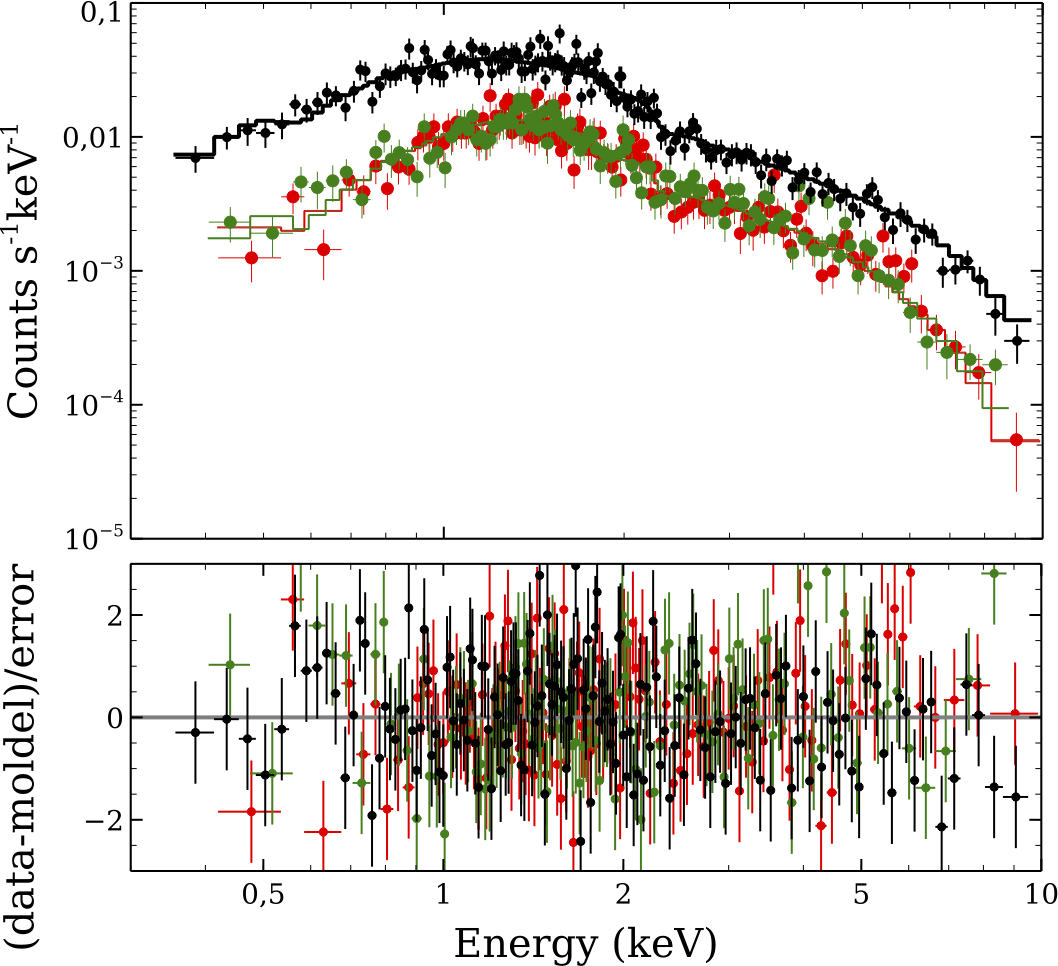}
\caption{Left, upper panel: Energy spectra of NGC 4861 X1 and X2 from C3 data. The black circles and solid black line represent X1 data and {\scshape pl} model, respectively. The red circles and solid red line represent X2 data and {\scshape dıskbb} model, respectively. {\scshape pl} model values shifted up factor of three for clarity. Right upper panel: {\scshape pl} energy spectra of NGC 4449 X7 from XM6 data. In {\it XMM-Newton} spectrum, black, green and red data points represent EPIC pn, MOS1 and MOS2, respectively. The model residuals are shown in the lower panels.}
\label{F:Fig4}
\end{center}
\end{figure*}

We analysed the unused published {\it Chandra} data (C2, C3, C4 and C5) to obtain spectral model parameters of NGC 4861 X1 and X2. The best-fit spectra were fitted with the {\scshape pl} and {\scshape dıskbb} models, together with the absorption model (tbabs). The X-ray absorption parameter was used as Galactic column density of $N_{H}$ = 1.1$\times10^{20}$ $cm^{-2}$ and this value was fixed. Unfortunately, the data are not good enough to constrain the local absorption. Therefore, only Galactic absorption value was used in fitting models to the spectral data of ULXs in NGC 4861.
The energy spectra of X1 and X2 using C3 data are given on the left panel in Fig. \ref{F:Fig4}.

For NGC 4449 X7, the acceptable {\scshape pl} model with two absorption (tbabs) models were obtained with XM6 data that was not used before. For this, the Galactic $N_{H}$ = 1.4$\times10^{20}$ $cm^{-2}$ value was taken from \citep{1990ARA&A..28..215D} and if this value is fixed, the second $N_{H}$ value can be kept as a free parameter. The energy spectrum of X7 is given in the right panel in Fig. \ref{F:Fig4} for the {\scshape pl} model.

The fluxes obtained by using the model with the absorption effect were calculated with {\it cflux} routine in the 0.3-10 keV energy range. The flux values and derived luminosities with adopted distances 9.95 Mpc for NGC 4861 and 4.02 Mpc for NGC 4449) are given in Table \ref{T:tab7} and Table \ref{T:tab8} for three ULXs. This way, we have reanalysed all the available X-ray data for these sources (see Table \ref{T:tab1}).

In order to investigate the long-term flux variations for these 3 ULXs we used  all available observations. However, for NGC 4861 X2, XM3 data was not included since its location corresponds to EPIC pn chip gap. The unabsorbed fluxes of ULXs were calculated in the energy range 0.3–10 keV using {\it cflux} in {\scshape xspec}. For X7 in NGC 4449, 
Any spectral model did not fit due to lack of statistics on {\it Swift}/XRT data (between 2007-2017). Therefore, the flux values of X7 for {\it Swift}/XRT data were calculated from {\it webPIMMS} \footnote{https://cxc.harvard.edu/toolkit/pimms.jsp} using count rates. Photon index of $\Gamma$=1.8 and  $N_{H}$ = 1.4$\times10^{20}$ $cm^{-2}$ were taken for this calculation.
 
\section{Results and Discussions}
\label{sec:res}

In this study, we investigated the optical candidates of three ULXs in two nearby (<10 Mpc) dwarf galaxies, NGC 4861 and NGC 4449. For this purpose, we used {\it HST} archival observations. We also analysed the data from {\it XMM-Newton}, {\it Chandra} and {\it Swift} to examine the X-ray properties of these sources. While 2018 datasets of two galaxies were analysed for the first time, the previous datasets were reanalysed.
As a result of the astrometric corrections, it was determined that optical counterpart of X1 corresponds to a dense HII region in NGC 4861 (Fig. \ref{Fig1}b). Therefore, it is not possible to resolve discrete source(s) as candidate(s) within the error radius of 0$\farcs$18. 

\begin{figure*}
	\begin{center}
		\includegraphics[width=\columnwidth]{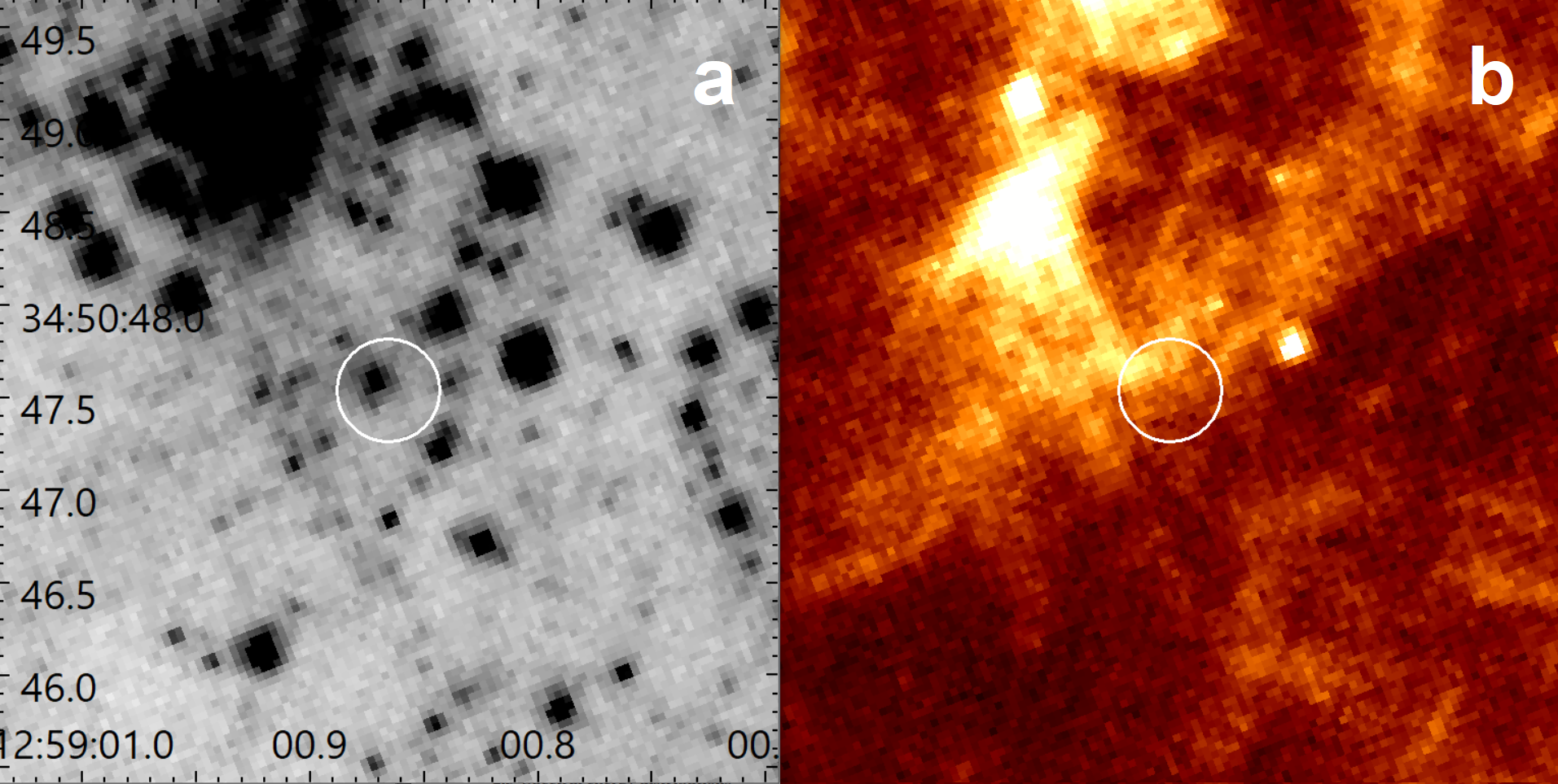}
		\caption{The F814W (a) and F658N (b) images of ULX X2 in NGC 4861 and its environment. White circles (0$\farcs$24 radius) show position of X2. The north is up. X-axis and y-axis represent R.A. and Dec., respectively.}
		\label{F:Fig5}
	\end{center}
\end{figure*}

In addition, for NGC 4861 X2, the other ULX, in this galaxy corresponds to an extended source within the error radius of 0$\farcs$24. Assuming that this source could be a star cluster we derived the age and mass values as 5 Myr and 400$\pm$80 M$\odot$, respectively. Regarding to its small mass, we considered this candidate object could be a small star group. The source X2 is located at the edge of intense HII region as seen in F658N image (Fig. \ref{F:Fig5}). The total H$_{\alpha}$ flux was calculated as 1.1$\times$10$^{-14}$ erg cm$^{-2}$ s$^{-1}$ from the 0$\farcs$24 circular region as given in Table \ref{T:tab4}. The excess of the flux contaminated by continuum emission within H$_{\alpha}$ bandwidth is clearly visible in this filter compared to others. To extract the nebular emission we applied the procedure as given by \cite{2011ApJ...734...33J}. Narrow band F658N image covers H$_{\alpha}$ and two adjacent lines such as [NII] $\lambda$6548 and [NII] $\lambda$6583. We built a normalized image from the weighted average of F814W and F547M. The resulting image was subtracted from the F658N image. The calculated nebular emission was determined as $\sim$ 70$\%$ of the total flux. Although this emission is high, it is difficult to associate with ULX X2 since the source is part of the adjacent to HII region.

Besides, three optical candidates were identified within the error radius of 0$\farcs$20 for NGC 4449 X7 from the {\it HST} images. {\it HST}/ACS F435W, F555W and F814W filters were used for the CMDs of the optical candidates with a nearby star cluster. The ages of the optical candidates were determined as 40 Myr for A1, A2 and 50 Myr for A3 and their masses were found to be $\sim$ 8 M$_{\odot}$ for all three of them. M$_v$ values of the optical candidates of X7 were determined as -5 mag for (A1 and A2) and -4.1 mag for A3. These values given in Table \ref{T:tab5} show that the candidates are within the range of known optical candidates of ULXs (-3 < M$_V$ < -8, \citealp{2015NatPh..11..551F, 2018ApJ...854..176V}). Recent studies have identified ULXs with absolute magnitudes Mv < -8, but such bright objects are extremely few \citep{2020gbar.conf...95V}. If the optical emission is thought to originate from donor star, spectral type of the candidates can be determined from their colors and absolute magnitude values. Using the Schmidt-Kaler table in \cite{1982lbg6.conf.....A}, the spectral type of A1, A2 and A3 were obtained as A, F and B type supergiants, respectively. Note that these donors are slightly more massive than the typical companions of ULXs with black holes obtained in some population calculations of X-ray binaries (main sequence star with mass of 6 M$_{\odot}$, \citealp{2017ApJ...846...17W}).

We have also investigated the properties of a star group very close to the NGC 4449 X7 which is not catalogued as a star cluster in \cite{2011ApJ...741...86R}. They have presented 129 compact and isolated star clusters brighter than M$_{V}\approx-7$ mag in NGC 4449 using HST observations. Here, the star group identified near X7 is not isolated and does not show a dense clustering. But we investigated this region within a circular area with a radius of 1$\farcs$0   to determine several physical parameters such as limit radius, mass and age.  This circular area is also marked as region 11 in the study of \citealp{2020ApJ...889..154W}. We constructed the radial density profile of this region by using King model \citep{1962AJ.....67..471K} to find the limit radius defined as the radius where the stellar density approaches background level.  For this, we counted stars in concentric rings of 0$\farcs$05 steps around the center and the limit radius is found to be 0.5$\arcsec$.

Unresolved bright regions in this limit radius (Fig. \ref{Fig2}b) correspond to young ($\sim$ 10 Myr) and bright (m$_{F555W}$ < 22 mag) sources in CMD. However, this age does not represent the true age of these sources since they are indistinguishable. On the other hand, considering the location of isolated and visually resolved stars in CMD, ages of stars in this small group are older than 40 Myr which is compatible with the ages of counterparts. Using SSP models and assuming all of the sources in the limited region to have the same age, we selected the two-part power-law IMF and found a total mass as $\sim$3$\times$10$^{4}$M$_{\odot}$. 
Although the distribution of stars in this region does not show a compact cluster structure as in the catalog of \cite{2011ApJ...741...86R}, their total brightness (M$_{V}=-8.7$ mag), age and mass values may indicate a cluster.

The X-ray properties of NGC 4861 X1 and X2 were previously investigated by \cite{2005ApJS..157...59L,2006MNRAS.370...25S,2009MNRAS.397.1061H,2014MNRAS.441.1841T}. In our study, we analysed all available X-ray data. For the NGC 4861 X1, we used the C2, C3, C4 and C5 data for the first time. After spectral modelling, both {\scshape pl} and {\scshape dıskpbb} models give acceptable fitting (see Table \ref{T:tab7}) due to the their C-statistics values except for C2 data.
For the {\scshape pl} model with photon indices $\Gamma$ $\sim$ (2.3-2.4), the unabsorbed luminosity, L$_{x}$ of NGC 4861 X1 falls into the range (2.5-2.8)$\times$10$^{39}$ erg s$^{-1}$. With this range, we interpret that the source luminosity does not vary in a relatively short time. When we reanalysed the older data sets (XM1, XM2 and XM3), we found that they were best fitted by {\scshape pl} model with photon indices ($\Gamma$ $\sim$ 1.9, 1.9 and 2.1). The calculated unabsorbed luminosity L$_{x}$, using the distance of 9.95 Mpc, lies within the range of (1.9-3.3)$\times$10$^{39}$ erg s$^{-1}$ in the 0.3-10 keV energy band. The spectrum of C1 is also adequately fitted by {\scshape pl} model ($\Gamma$ =1.9) with the highest unabsorbed luminosity of L$_{x}$=8.1$\times$10$^{39}$ erg s$^{-1}$. We see that the observed values of {\scshape pl} photon index for XM1, XM2 and C1 characterize the hard state (1.4<$\Gamma$<2.1) and for the remaining datasets the values of photon index are mostly on the low side for the steep power-law state ($\Gamma$ > 2.4) as described by \cite{2006ARA&A..44...49R}. It is known that
X-ray spectrum with a hard photon index is thought to originate from a corona. Also in the steep power-law state, the spectra could represent strong Comptonized thermal emission from an accretion disk of GBHBs and most of the them have higher luminosities in the steep power-law state than in the hard state. Although a similar correlation between luminosity and photon index has been found in some ULXs \citep{2006ApJ...650L..75F, 2009MNRAS.398.1450K, 2010MNRAS.407..291D}, we did not find such a correlation for the source X1. In fact, the luminosity of the source is clearly not correlated with $\Gamma$ to make firm conclusion.

Also, the spectra of NGC 4861 X1 were fitted by the {\scshape dıskpbb} model (called the {\it slim disk} model) allows the disk temperature to scale as T(r) $\propto$ r$^{-p}$, where r is the radius and {\it p} is a free parameter \citep{1994ApJ...426..308M}. If the standard model is expected, the value of {\it p} = 0.75 but in the case of an advection dominated disk {\it p} is between 0.5 and 0.75 \citep{2006PASJ...58..915V,2009MNRAS.397.1836G,2015ApJ...799..140S}. In our analysis, the spectral fits obtained show an acceptable range of C-statistic values with the exception of C2 data.
The derived values of spectral parameters appear to be in good agreement with the slim-disk model with {\it p} = 0.5, but for all cases, the p-values are not well constrained. Thereby, it is difficult to interpret from this model whether radial advection could take place and X1 is in a slim disk state with the derived inner-disk temperatures expected for super-Eddington accretion onto a stellar mass black hole.

As mentioned in several studies, when the data quality is low (e.g short exposure, distant galaxies), one-component models do not provide physically sufficient evidence to interpret the data, however these models should be taken into account statistically \citep{2009MNRAS.397.1836G,2013MNRAS.435.1758S,2017ARA&A..55..303K}. Therefore, we applied, {\scshape pl+dıskpbb} and {\scshape pl+dıskbb} models to available data but we did not achieve a significant improvement in the quality of  fits for the majority of the data. Although these models yield better statistics for some  datasets ( e.g. XM1 and XM2) model parameters are not constrained or they are not meaningful values as seen in Table \ref{T:tab7}. We conclude that, obtained data are insufficient to statistically distinguish simple models. Also, two-component models that describe high quality ULX spectra are not required for modelling.    

We obtained the best fitting single-component model parameters for NGC 4861 X2, given in Table \ref{T:tab7}. As seen from the results of spectral analysis of the data received within 10 days, it is not possible to distinguish between the {\scshape pl} and {\scshape dıskbb} models for the three out of four dataset according to their C-statistic values. However, for the C3 dataset, {\scshape dıskbb} model yields better fits than {\scshape pl} model on a 3-$\sigma$ confidence level according to F-test.
The temperature of {\scshape dıskbb} model was found as kT $\sim$ 0.80 keV. This value is comparable to the temperature $\sim$ 0.65 keV determined using XM1 and XM2 data. These kT values obtained are also consistent with the range of $\sim$ 0.7$-$1.5 keV for GBHBs with high mass accretion rates in their thermal state \citep{2006ARA&A..44...49R}.
The mass of the compact object in the ULX system can be estimated using the {\scshape dıskbb} model’s normalization parameter. For this calculation, the normalization value was obtained using C3 data and N$_{diskbb}$ = (r$_{in}$ / D$_{10}$)$^{2}$cos$\theta$ formula was used where r$_{in}$ is apparent inner disk radius from the observed data, D$_{10}$ is the distance in units of 10 kpc and $\theta$ is the angle of the disk. Here, the $\theta$ value was taken between 0$\degr$ - 75$\degr$ and we calculated r$_{in}$ $\sqrt{cosi}$ $\approx$ 90 - 170 km. True apparent radius (R$_{in}$) found between 105 - 205 km using the equation 3 in \cite{1998PASJ...50..667K}. Thus, the average mass of the compact object in X2 can be calculated as M $\sim$ 20 M$_{\odot}$ using the equation of M = R$_{in}$ / 8.86$\alpha$ M$_{\odot}$ given by \cite{2000ApJ...535..632M} where parameter $\alpha$ = 1 for a non-spinning and Schwarzschild black hole.

All archival data were reanalysed for NGC 4449 X7 then the best fitting spectral model parameters obtained for {\scshape pl} model. In 2001 C6 data we obtained a photon index of $\Gamma$= 1.97 and unabsorbed luminosity of L$_{x}$= 1.9$\times$10$^{39}$ erg s$^{-1}$, while in 2002 {\it XMM-Newton} data (XM4, XM5) $\Gamma$= 2.0$-$2.3 and L$_{x}$ is in the range of  (1.8$-$2.4)$\times$10$^{39}$ erg s$^{-1}$. Later, when looking at 2009 {\it Chandra} archive data (C7, C8) $\Gamma$= 2.6$-$2.9 and luminosity in the range of (0.9$-$1.7)$\times$10$^{39}$ erg s$^{-1}$ were obtained. The best model that fits the energy spectrum of X7 is {\scshape pl} ($\Gamma$=1.8) using XM6 data. The unabsorbed 0.3–10 keV luminosity of the source is L$_{x}$ $\approx$ 10$^{39}$ erg s$^{-1}$. In addition, the combined 2013 Swift-XRT spectrum is adequately fitted with the {\scshape pl} model with a photon index of $\Gamma$=1.5. Quite a similar unabsorbed luminosity value of L$_{x}$ =1.4$\times$10$^{39}$ erg s$^{-1}$ is also obtained.
These results show that the measured photon indices are in the range of (1.5-2.9) when the source X7 is represented by an absorbed {\scshape pl} model. $\Gamma$ < 2.1 values correspond to the low/hard state while steeper spectra (for C7 and C8) corresponding to the steep {\scshape pl} state. If we assume that the source exhibits possible spectral variation from hard state to steep {\scshape pl} state, this variation in luminosity seems not to correlate with the canonical GBHB states \citep{2006ARA&A..44...49R}.

Although two component models did not generally improve spectral results of X7, {\scshape pl+dıskbb} model gave the same C-statictics from {\scshape pl} model with XM6 data (see Table \ref{T:tab8}). This data used was based on a longer exposure time to provide sufficient counts for spectral fitting. Fraction of total flux in the soft model component ({\scshape diskbb}) is very low compare to hard ({\scshape pl}) model component over 0.3–10 keV band. Therefore, we could not have further analyses to investigate the spectral states of X7.

\begin{figure*}
\begin{center}
\includegraphics[width=\columnwidth]{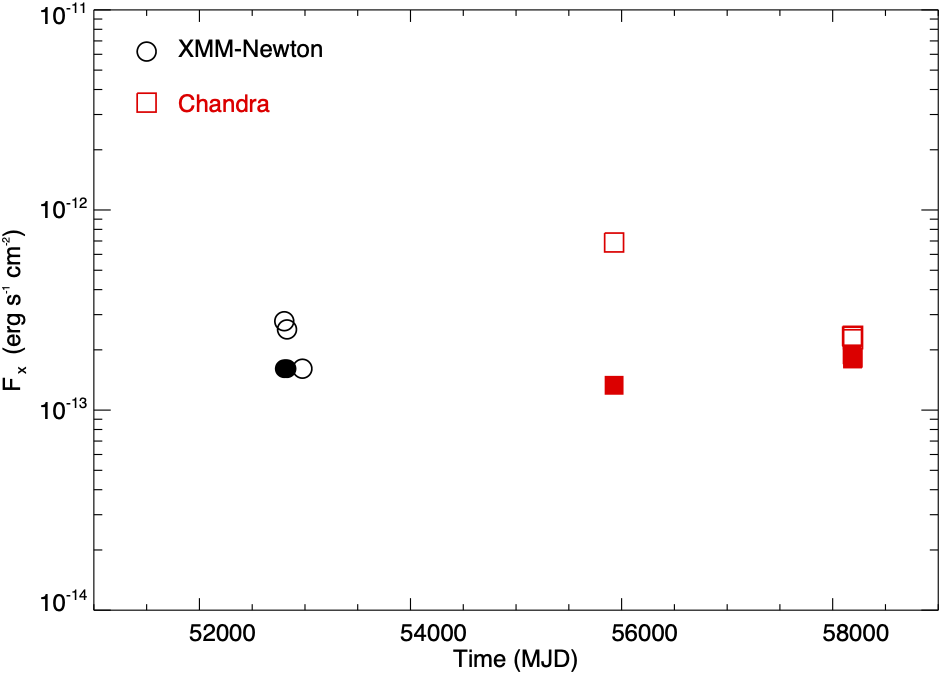}
\includegraphics[width=\columnwidth]{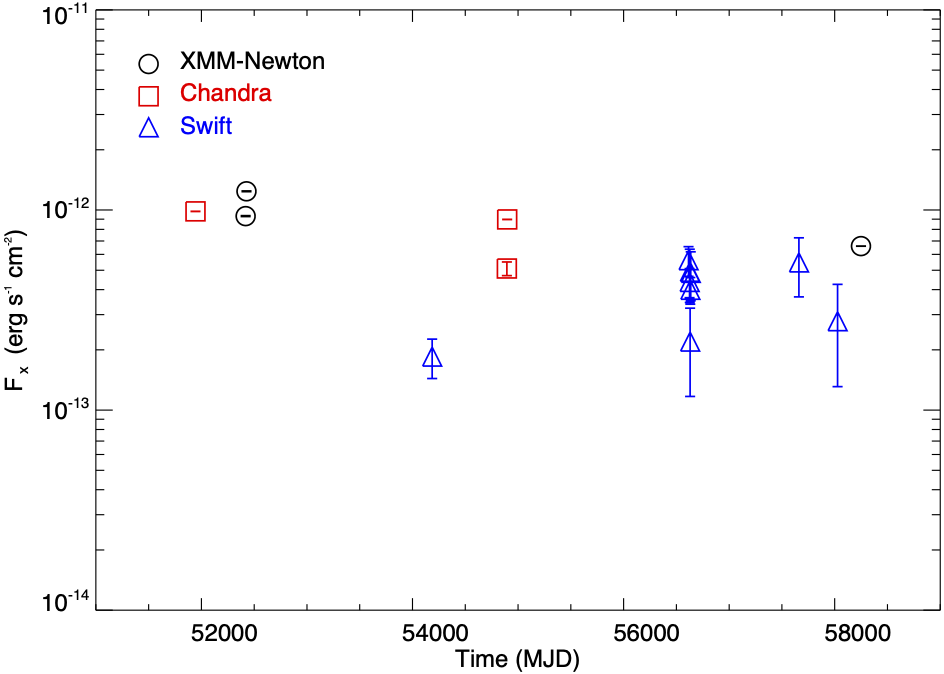}
\caption{The light curves of (left panel) NGC 4861 X1 (empty circles), X2 (filled circles) and (right panel) NGC 4449 X7 are shown. The black, red and blue colors represent {\it XMM-Newton}, {\it Chandra} and {\it Swift} data, respectively. Unabsorbed fluxes were obtained in the 0.3–10 keV energy band. In the left panel, error bars are not seen due to the very small error values.} 
\label{F:Fig7}
\end{center}
\end{figure*}

X-ray light curves in Fig. \ref{F:Fig7} show the flux variations between 2003 and 2018 for ULXs in NGC 4861. The flux values are varied as a factor of $\sim$ 4 for X1, with a minimum flux of 1.6$\times$10$^{-13}$ erg cm$^{-2}$ s$^{-1}$ in 2003 (XM3) and maximum flux of 6.9$\times$10$^{-13}$ erg cm$^{-2}$ s$^{-1}$ in 2012 (C1). On the other hand, the source X2 does not show any significant variability with the minimum value of 1.3$\times$10$^{-13}$ erg cm$^{-2}$ s$^{-1}$ in the 2012 C1 data and maximum flux value of 1.9$\times$10$^{-13}$ erg cm$^{-2}$ s$^{-1}$ in the C2 data from 2018. Our results are in line with the results of \cite{2014MNRAS.441.1841T} for these sources using more data. Also for NGC 4449 X7, a minimum flux is $\sim$ 2$\times$10$^{-13}$ erg cm$^{-2}$ s$^{-1}$ from {\it Swift} data and a maximum flux is $\sim$12$\times$10$^{-13}$ erg cm$^{-2}$ s$^{-1}$ from XM5 data implied a factor of $\sim$ 6 flux variation.

For many source classes such as AGN, BL Lac objects, galaxy clusters, normal galaxies, normal stars and X-ray pairs, the ratio of X-ray to optical fluxes ({\it{F$_x$}}/ {\it{F$_{opt}$}}) are determined to be in a different range for each source class \citep{1982ApJ...253..504M, 1991ApJS...76..813S}. This ratio satisfies log({\it{F$_x$}}/ {\it{F$_{opt}$}})=log({\it{F$_x$}})+m$_{V}$/2.5+5.37. Here, F$_{x}$ represents the unabsorbed X-ray flux in the 0.3$-$3.5 keV energy band and m$_{V}$ the extinction-corrected visual magnitude. Simultaneous X-ray and optical observations are not available for the two galaxies. Therefore, {\it XMM-Newton} (ID0112521701) archive data in 2002 and {\it HST} (j9cd01020) data set in 2005 were used for NGC 4449 X7. 
The {\it{F$_x$}}/ {\it{F$_{opt}$}} ratio was calculated in the range of 1.96 $-$ 2.30 for three optical candidates of X7.
These rates quite similar to other ULX's optical candidates \citep{2008ApJ...675.1067F,2011ApJ...737...81T,2011ApJ.733.118Y,2016ApJ...828..105A,2019MNRAS.488.5935A}.

We conclude that to constrain the spectral parameters better and to improve our understanding of the nature of these ULXs, further X-ray observations are needed. In addition, more sensitive photometric and spectroscopic observations should be performed in the optical band with the new generation telescopes.

\section*{Acknowledgements}
We thank the anonymous referee for providing very helpful comments.
We also thank M.E. Özel for his valuable contributions, comments and suggestions.
This research was supported by the Scientific and Technological Research Council of Turkey (TÜBİTAK) through project number 117F115. This research was also supported by the Çukurova University Research Fund through project number FBA-2019-11803 and funded by RFBR according to the research project 18-32-20214.

\section*{Data Availability}
The scientific results reported in this article are based on archival observations made by the {\it Chandra} X-ray Observatory, as well as archival observations by {\it XMM-Newton}, an ESA science mission with instruments and contributions directly funded by ESA Member States and NASA. This work has also made use of observations made with the NASA/ESA Hubble Space Telescope, and obtained from the data archive at the Space Telescope Science Institute. STScI is operated by the Association of Universities for Research in Astronomy, Inc. under NASA contract NAS 5-26555."

%%%%%%%%%%%%%%%%%%%% REFERENCES %%%%%%%%%%%%%%%%%%
% The best way to enter references is to use BibTeX:
\bibliographystyle{mnras}
\bibliography{ngc4449} % if your bibtex file is called 

\begin{thebibliography}{}
\makeatletter
\relax
\def\mn@urlcharsother{\let\do\@makeother \do\$\do\&\do\#\do\^\do\_\do\%\do\~}
\def\mn@doi{\begingroup\mn@urlcharsother \@ifnextchar [ {\mn@doi@}
  {\mn@doi@[]}}
\def\mn@doi@[#1]#2{\def\@tempa{#1}\ifx\@tempa\@empty \href
  {http://dx.doi.org/#2} {doi:#2}\else \href {http://dx.doi.org/#2} {#1}\fi
  \endgroup}
\def\mn@eprint#1#2{\mn@eprint@#1:#2::\@nil}
\def\mn@eprint@arXiv#1{\href {http://arxiv.org/abs/#1} {{\tt arXiv:#1}}}
\def\mn@eprint@dblp#1{\href {http://dblp.uni-trier.de/rec/bibtex/#1.xml}
  {dblp:#1}}
\def\mn@eprint@#1:#2:#3:#4\@nil{\def\@tempa {#1}\def\@tempb {#2}\def\@tempc
  {#3}\ifx \@tempc \@empty \let \@tempc \@tempb \let \@tempb \@tempa \fi \ifx
  \@tempb \@empty \def\@tempb {arXiv}\fi \@ifundefined
  {mn@eprint@\@tempb}{\@tempb:\@tempc}{\expandafter \expandafter \csname
  mn@eprint@\@tempb\endcsname \expandafter{\@tempc}}}

\bibitem[\protect\citeauthoryear{{Afanasiev} \& {Moiseev}}{{Afanasiev} \&
  {Moiseev}}{2005}]{2005AstL...31..194A}
{Afanasiev} V.~L.,  {Moiseev} A.~V.,  2005, \mn@doi [Astronomy Letters]
  {10.1134/1.1883351}, \href
  {https://ui.adsabs.harvard.edu/abs/2005AstL...31..194A} {31, 194}

\bibitem[\protect\citeauthoryear{{Aksaker}, {Akyuz}, {Avdan}  \&
  {Avdan}}{{Aksaker} et~al.}{2019}]{2019MNRAS.488.5935A}
{Aksaker} N.,  {Akyuz} A.,  {Avdan} S.,   {Avdan} H.,  2019, \mn@doi [\mnras]
  {10.1093/mnras/stz2160}, \href
  {https://ui.adsabs.harvard.edu/abs/2019MNRAS.488.5935A} {488, 5935}

\bibitem[\protect\citeauthoryear{{Aller} et~al.,}{{Aller}
  et~al.}{1982}]{1982lbg6.conf.....A}
{Aller} L.~H.,  et~al., eds, 1982, {Landolt-B{\"o}rnstein: Numerical Data and
  Functional Relationships in Science and Technology - New Series ``
  Gruppe/Group 6 Astronomy and Astrophysics '' Volume 2 Schaifers/Voigt:
  Astronomy and Astrophysics / Astronomie und Astrophysik `` Stars and Star
  Clusters / Sterne und Sternhaufen}

\bibitem[\protect\citeauthoryear{{Annibali}, {Aloisi}, {Mack}, {Tosi}, {van der
  Marel}, {Angeretti}, {Leitherer}  \& {Sirianni}}{{Annibali}
  et~al.}{2008}]{2008AJ....135.1900A}
{Annibali} F.,  {Aloisi} A.,  {Mack} J.,  {Tosi} M.,  {van der Marel} R.~P.,
  {Angeretti} L.,  {Leitherer} C.,   {Sirianni} M.,  2008, \mn@doi [\aj]
  {10.1088/0004-6256/135/5/1900}, \href
  {https://ui.adsabs.harvard.edu/abs/2008AJ....135.1900A} {135, 1900}

\bibitem[\protect\citeauthoryear{{Avdan}, {Avdan}, {Akyuz}, {Balman}, {Aksaker}
   \& {Akkaya Oralhan}}{{Avdan} et~al.}{2016}]{2016ApJ...828..105A}
{Avdan} H.,  {Avdan} S.,  {Akyuz} A.,  {Balman} S.,  {Aksaker} N.,   {Akkaya
  Oralhan} I.,  2016, \mn@doi [\apj] {10.3847/0004-637X/828/2/105}, \href
  {https://ui.adsabs.harvard.edu/abs/2016ApJ...828..105A} {828, 105}

\bibitem[\protect\citeauthoryear{{Avdan} et~al.,}{{Avdan}
  et~al.}{2019}]{2019ApJ...875...68A}
{Avdan} S.,  et~al., 2019, \mn@doi [\apj] {10.3847/1538-4357/ab10d9}, \href
  {https://ui.adsabs.harvard.edu/abs/2019ApJ...875...68A} {875, 68}

\bibitem[\protect\citeauthoryear{{Bachetti} et~al.,}{{Bachetti}
  et~al.}{2014}]{2014Natur.514..202B}
{Bachetti} M.,  et~al., 2014, \mn@doi [\nat] {10.1038/nature13791}, \href
  {http://adsabs.harvard.edu/abs/2014Natur.514..202B} {514, 202}

\bibitem[\protect\citeauthoryear{{Blair} \& {Fesen}}{{Blair} \&
  {Fesen}}{1998}]{1998AAS...193.7404B}
{Blair} W.~P.,  {Fesen} R.~A.,  1998, in American Astronomical Society Meeting
  Abstracts. p.~1365

\bibitem[\protect\citeauthoryear{{Bressan}, {Marigo}, {Girardi}, {Salasnich},
  {Dal Cero}, {Rubele}  \& {Nanni}}{{Bressan}
  et~al.}{2012}]{2012MNRAS.427..127B}
{Bressan} A.,  {Marigo} P.,  {Girardi} L.,  {Salasnich} B.,  {Dal Cero} C.,
  {Rubele} S.,   {Nanni} A.,  2012, \mn@doi [\mnras]
  {10.1111/j.1365-2966.2012.21948.x}, \href
  {https://ui.adsabs.harvard.edu/abs/2012MNRAS.427..127B} {427, 127}

\bibitem[\protect\citeauthoryear{{Bruzual} \& {Charlot}}{{Bruzual} \&
  {Charlot}}{2003}]{2003MNRAS.344..1000B}
{Bruzual} G.,  {Charlot} S.,  2003, \mn@doi [\mnras]
  {10.1046/j.1365-8711.2003.06897.x}, \href
  {http://www.iap.fr/~charlot/bc2003/paper} {344, 1000B}

\bibitem[\protect\citeauthoryear{{Carpano}, {Haberl}, {Maitra}  \&
  {Vasilopoulos}}{{Carpano} et~al.}{2018}]{2018MNRAS.476L..45C}
{Carpano} S.,  {Haberl} F.,  {Maitra} C.,   {Vasilopoulos} G.,  2018, \mn@doi
  [\mnras] {10.1093/mnrasl/sly030}, \href
  {https://ui.adsabs.harvard.edu/abs/2018MNRAS.476L..45C} {476, L45}

\bibitem[\protect\citeauthoryear{{Cash}}{{Cash}}{1979}]{1979ApJ...228..939C}
{Cash} W.,  1979, \mn@doi [\apj] {10.1086/156922}, \href
  {https://ui.adsabs.harvard.edu/abs/1979ApJ...228..939C} {228, 939}

\bibitem[\protect\citeauthoryear{{Colbert} \& {Ptak}}{{Colbert} \&
  {Ptak}}{2002}]{2002ApJS..143...25C}
{Colbert} E.~J.~M.,  {Ptak} A.~F.,  2002, \mn@doi [\apjs] {10.1086/342507},
  \href {https://ui.adsabs.harvard.edu/abs/2002ApJS..143...25C} {143, 25}

\bibitem[\protect\citeauthoryear{{Cormier} et~al.,}{{Cormier}
  et~al.}{2014}]{2014A&A...564A.121C}
{Cormier} D.,  et~al., 2014, \mn@doi [\aap] {10.1051/0004-6361/201322096},
  \href {https://ui.adsabs.harvard.edu/abs/2014A&A...564A.121C} {564, A121}

\bibitem[\protect\citeauthoryear{{Dewangan}, {Misra}, {Rao}  \&
  {Griffiths}}{{Dewangan} et~al.}{2010}]{2010MNRAS.407..291D}
{Dewangan} G.~C.,  {Misra} R.,  {Rao} A.~R.,   {Griffiths} R.~E.,  2010,
  \mn@doi [\mnras] {10.1111/j.1365-2966.2010.16893.x}, \href
  {https://ui.adsabs.harvard.edu/abs/2010MNRAS.407..291D} {407, 291}

\bibitem[\protect\citeauthoryear{{Dickey} \& {Lockman}}{{Dickey} \&
  {Lockman}}{1990}]{1990ARA&A..28..215D}
{Dickey} J.~M.,  {Lockman} F.~J.,  1990, \mn@doi [\araa]
  {10.1146/annurev.aa.28.090190.001243}, \href
  {https://ui.adsabs.harvard.edu/abs/1990ARA%26A..28..215D} {28, 215}

\bibitem[\protect\citeauthoryear{{Dolphin}}{{Dolphin}}{2000}]{2000PASP..112.1383D}
{Dolphin} A.~E.,  2000, \mn@doi [\pasp] {10.1086/316630}, \href
  {http://adsabs.harvard.edu/abs/2000PASP..112.1383D} {112, 1383}

\bibitem[\protect\citeauthoryear{{Egorov}, {Lozinskaya}  \& {Moiseev}}{{Egorov}
  et~al.}{2017}]{2017MNRAS.467L...1E}
{Egorov} O.~V.,  {Lozinskaya} T.~A.,   {Moiseev} A.~V.,  2017, \mn@doi [\mnras]
  {10.1093/mnrasl/slw246}, \href
  {https://ui.adsabs.harvard.edu/abs/2017MNRAS.467L...1E} {467, L1}

\bibitem[\protect\citeauthoryear{{Esposito}, {Israel}, {Milisavljevic},
  {Mapelli}, {Zampieri}, {Sidoli}, {Fabbiano}  \& {Rodr{\'\i}guez
  Castillo}}{{Esposito} et~al.}{2015}]{2015MNRAS.452.1112E}
{Esposito} P.,  {Israel} G.~L.,  {Milisavljevic} D.,  {Mapelli} M.,  {Zampieri}
  L.,  {Sidoli} L.,  {Fabbiano} G.,   {Rodr{\'\i}guez Castillo} G.~A.,  2015,
  \mn@doi [\mnras] {10.1093/mnras/stv1379}, \href
  {https://ui.adsabs.harvard.edu/abs/2015MNRAS.452.1112E} {452, 1112}

\bibitem[\protect\citeauthoryear{{Esteban}, {Bresolin}, {Peimbert},
  {Garc{\'\i}a-Rojas}, {Peimbert}  \& {Mesa-Delgado}}{{Esteban}
  et~al.}{2009}]{2009ApJ...700..654E}
{Esteban} C.,  {Bresolin} F.,  {Peimbert} M.,  {Garc{\'\i}a-Rojas} J.,
  {Peimbert} A.,   {Mesa-Delgado} A.,  2009, \mn@doi [\apj]
  {10.1088/0004-637X/700/1/654}, \href
  {https://ui.adsabs.harvard.edu/abs/2009ApJ...700..654E} {700, 654}

\bibitem[\protect\citeauthoryear{{Fabrika}, {Ueda}, {Vinokurov}, {Sholukhova}
  \& {Shidatsu}}{{Fabrika} et~al.}{2015}]{2015NatPh..11..551F}
{Fabrika} S.,  {Ueda} Y.,  {Vinokurov} A.,  {Sholukhova} O.,   {Shidatsu} M.,
  2015, \mn@doi [Nature Physics] {10.1038/nphys3348}, \href
  {https://ui.adsabs.harvard.edu/abs/2015NatPh..11..551F} {11, 551}

\bibitem[\protect\citeauthoryear{{Farrell}, {Webb}, {Barret}, {Godet}  \&
  {Rodrigues}}{{Farrell} et~al.}{2009}]{2009Natur.460...73F}
{Farrell} S.~A.,  {Webb} N.~A.,  {Barret} D.,  {Godet} O.,   {Rodrigues} J.~M.,
   2009, \mn@doi [\nat] {10.1038/nature08083}, \href
  {http://adsabs.harvard.edu/abs/2009Natur.460...73F} {460, 73}

\bibitem[\protect\citeauthoryear{{Feng} \& {Kaaret}}{{Feng} \&
  {Kaaret}}{2005}]{2005ApJ...633.1052F}
{Feng} H.,  {Kaaret} P.,  2005, \mn@doi [\apj] {10.1086/491597}, \href
  {https://ui.adsabs.harvard.edu/abs/2005ApJ...633.1052F} {633, 1052}

\bibitem[\protect\citeauthoryear{{Feng} \& {Kaaret}}{{Feng} \&
  {Kaaret}}{2006}]{2006ApJ...650L..75F}
{Feng} H.,  {Kaaret} P.,  2006, \mn@doi [\apjl] {10.1086/508613}, \href
  {https://ui.adsabs.harvard.edu/abs/2006ApJ...650L..75F} {650, L75}

\bibitem[\protect\citeauthoryear{{Feng} \& {Kaaret}}{{Feng} \&
  {Kaaret}}{2008}]{2008ApJ...675.1067F}
{Feng} H.,  {Kaaret} P.,  2008, \mn@doi [\apj] {10.1086/527353}, \href
  {http://adsabs.harvard.edu/abs/2008ApJ...675.1067F} {675, 1067}

\bibitem[\protect\citeauthoryear{{French}}{{French}}{1980}]{1980ApJ...240...41F}
{French} H.~B.,  1980, \mn@doi [\apj] {10.1086/158205}, \href
  {https://ui.adsabs.harvard.edu/abs/1980ApJ...240...41F} {240, 41}

\bibitem[\protect\citeauthoryear{{F{\"u}rst} et~al.,}{{F{\"u}rst}
  et~al.}{2016}]{2016ApJ...831L..14F}
{F{\"u}rst} F.,  et~al., 2016, \mn@doi [\apjl] {10.3847/2041-8205/831/2/L14},
  \href {https://ui.adsabs.harvard.edu/abs/2016ApJ...831L..14F} {831, L14}

\bibitem[\protect\citeauthoryear{{Gaia Collaboration} et~al.,}{{Gaia
  Collaboration} et~al.}{2018}]{2018A&A...616A...1G}
{Gaia Collaboration} et~al., 2018, \mn@doi [Astronomy and Astrophysics]
  {10.1051/0004-6361/201833051}, \href
  {https://ui.adsabs.harvard.edu/abs/2018A&A...616A...1G} {616, A1}

\bibitem[\protect\citeauthoryear{{Gladstone}, {Roberts}  \& {Done}}{{Gladstone}
  et~al.}{2009}]{2009MNRAS.397.1836G}
{Gladstone} J.~C.,  {Roberts} T.~P.,   {Done} C.,  2009, \mn@doi [\mnras]
  {10.1111/j.1365-2966.2009.15123.x}, \href
  {https://ui.adsabs.harvard.edu/abs/2009MNRAS.397.1836G} {397, 1836}

\bibitem[\protect\citeauthoryear{{G{\"o}tberg}, {de Mink}, {Groh}, {Leitherer}
  \& {Norman}}{{G{\"o}tberg} et~al.}{2019}]{2019A&A...629A.134G}
{G{\"o}tberg} Y.,  {de Mink} S.~E.,  {Groh} J.~H.,  {Leitherer} C.,   {Norman}
  C.,  2019, \mn@doi [\aap] {10.1051/0004-6361/201834525}, \href
  {https://ui.adsabs.harvard.edu/abs/2019A&A...629A.134G} {629, A134}

\bibitem[\protect\citeauthoryear{{Gris{\'e}}, {Pakull}  \& {Motch}}{{Gris{\'e}}
  et~al.}{2005}]{2005sf2a.conf..549G}
{Gris{\'e}} F.,  {Pakull} M.,   {Motch} C.,  2005, in {Casoli} F.,  {Contini}
  T.,  {Hameury} J.~M.,   {Pagani} L.,  eds, SF2A-2005: Semaine de
  l'Astrophysique Francaise. p.~549

\bibitem[\protect\citeauthoryear{{Gris{\'e}}, {Kaaret}, {Pakull}  \&
  {Motch}}{{Gris{\'e}} et~al.}{2011}]{2011ApJ...734...23G}
{Gris{\'e}} F.,  {Kaaret} P.,  {Pakull} M.~W.,   {Motch} C.,  2011, \mn@doi
  [\apj] {10.1088/0004-637X/734/1/23}, \href
  {http://adsabs.harvard.edu/abs/2011ApJ...734...23G} {734, 23}

\bibitem[\protect\citeauthoryear{{Gris{\'e}}, {Kaaret}, {Corbel}, {Feng},
  {Cseh}  \& {Tao}}{{Gris{\'e}} et~al.}{2012}]{2012ApJ...745..123G}
{Gris{\'e}} F.,  {Kaaret} P.,  {Corbel} S.,  {Feng} H.,  {Cseh} D.,   {Tao} L.,
   2012, \mn@doi [\apj] {10.1088/0004-637X/745/2/123}, \href
  {https://ui.adsabs.harvard.edu/abs/2012ApJ...745..123G} {745, 123}

\bibitem[\protect\citeauthoryear{{Heida} et~al.,}{{Heida}
  et~al.}{2015}]{2015MNRAS.453.3510H}
{Heida} M.,  et~al., 2015, \mn@doi [\mnras] {10.1093/mnras/stv1853}, \href
  {https://ui.adsabs.harvard.edu/abs/2015MNRAS.453.3510H} {453, 3510}

\bibitem[\protect\citeauthoryear{{Heida}, {Jonker}, {Torres}, {Roberts},
  {Walton}, {Moon}, {Stern}  \& {Harrison}}{{Heida}
  et~al.}{2016}]{2016MNRAS.459..771H}
{Heida} M.,  {Jonker} P.~G.,  {Torres} M.~A.~P.,  {Roberts} T.~P.,  {Walton}
  D.~J.,  {Moon} D.~S.,  {Stern} D.,   {Harrison} F.~A.,  2016, \mn@doi
  [\mnras] {10.1093/mnras/stw695}, \href
  {https://ui.adsabs.harvard.edu/abs/2016MNRAS.459..771H} {459, 771}

\bibitem[\protect\citeauthoryear{{Heida} et~al.,}{{Heida}
  et~al.}{2019}]{2019ApJ...883L..34H}
{Heida} M.,  et~al., 2019, \mn@doi [\apjl] {10.3847/2041-8213/ab4139}, \href
  {https://ui.adsabs.harvard.edu/abs/2019ApJ...883L..34H} {883, L34}

\bibitem[\protect\citeauthoryear{{Heil}, {Vaughan}  \& {Roberts}}{{Heil}
  et~al.}{2009}]{2009MNRAS.397.1061H}
{Heil} L.~M.,  {Vaughan} S.,   {Roberts} T.~P.,  2009, \mn@doi [\mnras]
  {10.1111/j.1365-2966.2009.15068.x}, \href
  {https://ui.adsabs.harvard.edu/abs/2009MNRAS.397.1061H} {397, 1061}

\bibitem[\protect\citeauthoryear{{Hill} et~al.,}{{Hill}
  et~al.}{1998}]{1998ApJ...507..179H}
{Hill} R.~S.,  et~al., 1998, \mn@doi [\apj] {10.1086/306302}, \href
  {https://ui.adsabs.harvard.edu/abs/1998ApJ...507..179H} {507, 179}

\bibitem[\protect\citeauthoryear{{Israel} et~al.,}{{Israel}
  et~al.}{2017a}]{2017Sci...355..817I}
{Israel} G.~L.,  et~al., 2017a, \mn@doi [Science] {10.1126/science.aai8635},
  \href {https://ui.adsabs.harvard.edu/abs/2017Sci...355..817I} {355, 817}

\bibitem[\protect\citeauthoryear{{Israel} et~al.,}{{Israel}
  et~al.}{2017b}]{2017MNRAS.466L..48I}
{Israel} G.~L.,  et~al., 2017b, \mn@doi [\mnras] {10.1093/mnrasl/slw218}, \href
  {https://ui.adsabs.harvard.edu/abs/2017MNRAS.466L..48I} {466, L48}

\bibitem[\protect\citeauthoryear{{Izotov} \& {Thuan}}{{Izotov} \&
  {Thuan}}{1999}]{1999ApJ...511..639I}
{Izotov} Y.~I.,  {Thuan} T.~X.,  1999, \mn@doi [\apj] {10.1086/306708}, \href
  {https://ui.adsabs.harvard.edu/abs/1999ApJ...511..639I} {511, 639}

\bibitem[\protect\citeauthoryear{{Izotov}, {Schaerer}, {Blecha}, {Royer},
  {Guseva}  \& {North}}{{Izotov} et~al.}{2006}]{2006A&A...459...71I}
{Izotov} Y.~I.,  {Schaerer} D.,  {Blecha} A.,  {Royer} F.,  {Guseva} N.~G.,
  {North} P.,  2006, \mn@doi [\aap] {10.1051/0004-6361:20065622}, \href
  {https://ui.adsabs.harvard.edu/abs/2006A&A...459...71I} {459, 71}

\bibitem[\protect\citeauthoryear{{Jenkins} et~al.,}{{Jenkins}
  et~al.}{2011}]{2011ApJ...734...33J}
{Jenkins} L.~P.,  et~al., 2011, \mn@doi [\apj] {10.1088/0004-637X/734/1/33},
  \href {https://ui.adsabs.harvard.edu/abs/2011ApJ...734...33J} {734, 33}

\bibitem[\protect\citeauthoryear{{Jonker} et~al.,}{{Jonker}
  et~al.}{2012}]{2012ApJ...758...28J}
{Jonker} P.~G.,  et~al., 2012, \mn@doi [\apj] {10.1088/0004-637X/758/1/28},
  \href {https://ui.adsabs.harvard.edu/abs/2012ApJ...758...28J} {758, 28}

\bibitem[\protect\citeauthoryear{{Kaaret}, {Feng}  \& {Roberts}}{{Kaaret}
  et~al.}{2017}]{2017ARA&A..55..303K}
{Kaaret} P.,  {Feng} H.,   {Roberts} T.~P.,  2017, \mn@doi [\araa]
  {10.1146/annurev-astro-091916-055259}, \href
  {http://adsabs.harvard.edu/abs/2017ARA%26A..55..303K} {55, 303}

\bibitem[\protect\citeauthoryear{{Kajava} \& {Poutanen}}{{Kajava} \&
  {Poutanen}}{2009}]{2009MNRAS.398.1450K}
{Kajava} J. J.~E.,  {Poutanen} J.,  2009, \mn@doi [\mnras]
  {10.1111/j.1365-2966.2009.15215.x}, \href
  {https://ui.adsabs.harvard.edu/abs/2009MNRAS.398.1450K} {398, 1450}

\bibitem[\protect\citeauthoryear{{King}}{{King}}{1962}]{1962AJ.....67..471K}
{King} I.,  1962, \mn@doi [\aj] {10.1086/108756}, \href
  {https://ui.adsabs.harvard.edu/abs/1962AJ.....67..471K} {67, 471}

\bibitem[\protect\citeauthoryear{{Kobulnicky} \& {Skillman}}{{Kobulnicky} \&
  {Skillman}}{1998}]{1998ApJ...497..601K}
{Kobulnicky} H.~A.,  {Skillman} E.~D.,  1998, \mn@doi [\apj] {10.1086/305491},
  \href {https://ui.adsabs.harvard.edu/abs/1998ApJ...497..601K} {497, 601}

\bibitem[\protect\citeauthoryear{{Kroupa}}{{Kroupa}}{2001}]{2001MNRAS.322..231K}
{Kroupa} P.,  2001, \mn@doi [\mnras] {10.1046/j.1365-8711.2001.04022.x}, \href
  {http://articles.adsabs.harvard.edu/pdf/2001MNRAS.322..231K} {322, 231}

\bibitem[\protect\citeauthoryear{{Kubota}, {Tanaka}, {Makishima}, {Ueda},
  {Dotani}, {Inoue}  \& {Yamaoka}}{{Kubota} et~al.}{1998}]{1998PASJ...50..667K}
{Kubota} A.,  {Tanaka} Y.,  {Makishima} K.,  {Ueda} Y.,  {Dotani} T.,  {Inoue}
  H.,   {Yamaoka} K.,  1998, \mn@doi [\pasj] {10.1093/pasj/50.6.667}, \href
  {https://ui.adsabs.harvard.edu/abs/1998PASJ...50..667K} {50, 667}

\bibitem[\protect\citeauthoryear{{Leitherer} et~al.,}{{Leitherer}
  et~al.}{1999}]{1999ApJS..123....3L}
{Leitherer} C.,  et~al., 1999, \mn@doi [\apjs] {10.1086/313233}, \href
  {https://ui.adsabs.harvard.edu/abs/1999ApJS..123....3L} {123, 3}

\bibitem[\protect\citeauthoryear{{Liu} \& {Bregman}}{{Liu} \&
  {Bregman}}{2005}]{2005ApJS..157...59L}
{Liu} J.-F.,  {Bregman} J.~N.,  2005, \mn@doi [\apjs] {10.1086/427170}, \href
  {https://ui.adsabs.harvard.edu/abs/2005ApJS..157...59L} {157, 59}

\bibitem[\protect\citeauthoryear{{Liu}, {Bregman}  \& {Seitzer}}{{Liu}
  et~al.}{2004}]{2004ApJ...602..249L}
{Liu} J.-F.,  {Bregman} J.~N.,   {Seitzer} P.,  2004, \mn@doi [\apj]
  {10.1086/380994}, \href {http://adsabs.harvard.edu/abs/2004ApJ...602..249L}
  {602, 249}

\bibitem[\protect\citeauthoryear{{Liu}, {Bregman}, {Bai}, {Justham}  \&
  {Crowther}}{{Liu} et~al.}{2013}]{2013Natur.503..500L}
{Liu} J.-F.,  {Bregman} J.~N.,  {Bai} Y.,  {Justham} S.,   {Crowther} P.,
  2013, \mn@doi [\nat] {10.1038/nature12762}, \href
  {https://ui.adsabs.harvard.edu/abs/2013Natur.503..500L} {503, 500}

\bibitem[\protect\citeauthoryear{{Loose} \& {Thuan}}{{Loose} \&
  {Thuan}}{1986}]{1986ApJ...309...59L}
{Loose} H.-H.,  {Thuan} T.~X.,  1986, \mn@doi [\apj] {10.1086/164577}, \href
  {https://ui.adsabs.harvard.edu/abs/1986ApJ...309...59L} {309, 59}

\bibitem[\protect\citeauthoryear{{L{\'o}pez}, {Heida}, {Jonker}, {Torres},
  {Roberts}, {Walton}, {Moon}  \& {Harrison}}{{L{\'o}pez}
  et~al.}{2017}]{2017MNRAS.469..671L}
{L{\'o}pez} K.~M.,  {Heida} M.,  {Jonker} P.~G.,  {Torres} M.~A.~P.,  {Roberts}
  T.~P.,  {Walton} D.~J.,  {Moon} D.~S.,   {Harrison} F.~A.,  2017, \mn@doi
  [\mnras] {10.1093/mnras/stx857}, \href
  {https://ui.adsabs.harvard.edu/abs/2017MNRAS.469..671L} {469, 671}

\bibitem[\protect\citeauthoryear{{Maccacaro} et~al.,}{{Maccacaro}
  et~al.}{1982}]{1982ApJ...253..504M}
{Maccacaro} T.,  et~al., 1982, \mn@doi [\apj] {10.1086/159652}, \href
  {https://ui.adsabs.harvard.edu/abs/1982ApJ...253..504M} {253, 504}

\bibitem[\protect\citeauthoryear{{Makishima} et~al.,}{{Makishima}
  et~al.}{2000}]{2000ApJ...535..632M}
{Makishima} K.,  et~al., 2000, \mn@doi [\apj] {10.1086/308868}, \href
  {http://adsabs.harvard.edu/abs/2000ApJ...535..632M} {535, 632}

\bibitem[\protect\citeauthoryear{{Mezcua}, {Lobanov}  \&
  {Mart{\'\i}-Vidal}}{{Mezcua} et~al.}{2013}]{2013MNRAS.436.2454M}
{Mezcua} M.,  {Lobanov} A.~P.,   {Mart{\'\i}-Vidal} I.,  2013, \mn@doi [Monthly
  Notices of the Royal Astronomical Society] {10.1093/mnras/stt1738}, \href
  {https://ui.adsabs.harvard.edu/abs/2013MNRAS.436.2454M} {436, 2454}

\bibitem[\protect\citeauthoryear{{Mezcua}, {Roberts}, {Lobanov}  \&
  {Sutton}}{{Mezcua} et~al.}{2015}]{2015MNRAS.448.1893M}
{Mezcua} M.,  {Roberts} T.~P.,  {Lobanov} A.~P.,   {Sutton} A.~D.,  2015,
  \mn@doi [\mnras] {10.1093/mnras/stv143}, 448, 1893

\bibitem[\protect\citeauthoryear{{Miller} \& {Colbert}}{{Miller} \&
  {Colbert}}{2004}]{2004IJMPD..13....1M}
{Miller} M.~C.,  {Colbert} E.~J.~M.,  2004, \mn@doi [International Journal of
  Modern Physics D] {10.1142/S0218271804004426}, \href
  {https://ui.adsabs.harvard.edu/abs/2004IJMPD..13....1M} {13, 1}

\bibitem[\protect\citeauthoryear{{Mineshige}, {Hirano}, {Kitamoto}, {Yamada}
  \& {Fukue}}{{Mineshige} et~al.}{1994}]{1994ApJ...426..308M}
{Mineshige} S.,  {Hirano} A.,  {Kitamoto} S.,  {Yamada} T.~T.,   {Fukue} J.,
  1994, \mn@doi [\apj] {10.1086/174065}, \href
  {https://ui.adsabs.harvard.edu/abs/1994ApJ...426..308M} {426, 308}

\bibitem[\protect\citeauthoryear{{Motch}, {Pakull}, {Gris{\'e}}  \&
  {Soria}}{{Motch} et~al.}{2011}]{2011AN....332..367M}
{Motch} C.,  {Pakull} M.~W.,  {Gris{\'e}} F.,   {Soria} R.,  2011, \mn@doi
  [Astronomische Nachrichten] {10.1002/asna.201011501}, \href
  {https://ui.adsabs.harvard.edu/abs/2011AN....332..367M} {332, 367}

\bibitem[\protect\citeauthoryear{{Motch}, {Pakull}, {Soria}, {Gris{\'e}}  \&
  {Pietrzy{\'n}ski}}{{Motch} et~al.}{2014}]{2014Natur.514..198M}
{Motch} C.,  {Pakull} M.~W.,  {Soria} R.,  {Gris{\'e}} F.,   {Pietrzy{\'n}ski}
  G.,  2014, \mn@doi [\nat] {10.1038/nature13730}, \href
  {https://ui.adsabs.harvard.edu/abs/2014Natur.514..198M} {514, 198}

\bibitem[\protect\citeauthoryear{{Patruno} \& {Zampieri}}{{Patruno} \&
  {Zampieri}}{2010}]{2010MNRAS.403L..69P}
{Patruno} A.,  {Zampieri} L.,  2010, \mn@doi [\mnras]
  {10.1111/j.1745-3933.2010.00817.x}, \href
  {https://ui.adsabs.harvard.edu/abs/2010MNRAS.403L..69P} {403, L69}

\bibitem[\protect\citeauthoryear{{Poutanen}, {Lipunova}, {Fabrika}, {Butkevich}
   \& {Abolmasov}}{{Poutanen} et~al.}{2007}]{2007MNRAS.377.1187P}
{Poutanen} J.,  {Lipunova} G.,  {Fabrika} S.,  {Butkevich} A.~G.,   {Abolmasov}
  P.,  2007, \mn@doi [\mnras] {10.1111/j.1365-2966.2007.11668.x}, \href
  {https://ui.adsabs.harvard.edu/abs/2007MNRAS.377.1187P} {377, 1187}

\bibitem[\protect\citeauthoryear{{Poutanen}, {Fabrika}, {Valeev}, {Sholukhova}
  \& {Greiner}}{{Poutanen} et~al.}{2013}]{2013MNRAS.432..506P}
{Poutanen} J.,  {Fabrika} S.,  {Valeev} A.~F.,  {Sholukhova} O.,   {Greiner}
  J.,  2013, \mn@doi [\mnras] {10.1093/mnras/stt487}, \href
  {https://ui.adsabs.harvard.edu/abs/2013MNRAS.432..506P} {432, 506}

\bibitem[\protect\citeauthoryear{{Qiu} et~al.,}{{Qiu}
  et~al.}{2019}]{2019ApJ...877...57Q}
{Qiu} Y.,  et~al., 2019, \mn@doi [\apj] {10.3847/1538-4357/ab16e7}, \href
  {https://ui.adsabs.harvard.edu/abs/2019ApJ...877...57Q} {877, 57}

\bibitem[\protect\citeauthoryear{{Rangelov}, {Prestwich}  \&
  {Chandar}}{{Rangelov} et~al.}{2011}]{2011ApJ...741...86R}
{Rangelov} B.,  {Prestwich} A.~H.,   {Chandar} R.,  2011, \mn@doi [\apj]
  {10.1088/0004-637X/741/2/86}, \href
  {https://ui.adsabs.harvard.edu/abs/2011ApJ...741...86R} {741, 86}

\bibitem[\protect\citeauthoryear{{Remillard} \& {McClintock}}{{Remillard} \&
  {McClintock}}{2006}]{2006ARA&A..44...49R}
{Remillard} R.~A.,  {McClintock} J.~E.,  2006, \mn@doi [\araa]
  {10.1146/annurev.astro.44.051905.092532}, \href
  {https://ui.adsabs.harvard.edu/abs/2006ARA&A..44...49R} {44, 49}

\bibitem[\protect\citeauthoryear{{Roberts}, {Levan}  \& {Goad}}{{Roberts}
  et~al.}{2008}]{2008MNRAS.387...73R}
{Roberts} T.~P.,  {Levan} A.~J.,   {Goad} M.~R.,  2008, \mn@doi [\mnras]
  {10.1111/j.1365-2966.2008.13293.x}, \href
  {https://ui.adsabs.harvard.edu/abs/2008MNRAS.387...73R} {387, 73}

\bibitem[\protect\citeauthoryear{{Rodr{\'\i}guez Castillo}
  et~al.,}{{Rodr{\'\i}guez Castillo} et~al.}{2020}]{2020ApJ...895...60R}
{Rodr{\'\i}guez Castillo} G.~A.,  et~al., 2020, \mn@doi [\apj]
  {10.3847/1538-4357/ab8a44}, \href
  {https://ui.adsabs.harvard.edu/abs/2020ApJ...895...60R} {895, 60}

\bibitem[\protect\citeauthoryear{{Sabbi} et~al.,}{{Sabbi}
  et~al.}{2018}]{2018ApJS..235...23S}
{Sabbi} E.,  et~al., 2018, \mn@doi [\apjs] {10.3847/1538-4365/aaa8e5}, \href
  {https://ui.adsabs.harvard.edu/abs/2018ApJS..235...23S} {235, 23}

\bibitem[\protect\citeauthoryear{{Sarkisyan}, {Vinokurov}, {Solovieva},
  {Sholukhova}, {Kostenkov}  \& {Fabrika}}{{Sarkisyan}
  et~al.}{2017}]{2017AstBu..72..486S}
{Sarkisyan} A.~N.,  {Vinokurov} A.~S.,  {Solovieva} Y.~N.,  {Sholukhova} O.~N.,
   {Kostenkov} A.~E.,   {Fabrika} S.~N.,  2017, \mn@doi [Astrophysical
  Bulletin] {10.1134/S1990341317040137}, \href
  {https://ui.adsabs.harvard.edu/abs/2017AstBu..72..486S} {72, 486}

\bibitem[\protect\citeauthoryear{{Sathyaprakash} et~al.,}{{Sathyaprakash}
  et~al.}{2019}]{2019MNRAS.488L..35S}
{Sathyaprakash} R.,  et~al., 2019, \mn@doi [\mnras] {10.1093/mnrasl/slz086},
  \href {https://ui.adsabs.harvard.edu/abs/2019MNRAS.488L..35S} {488, L35}

\bibitem[\protect\citeauthoryear{{Schlafly} \& {Finkbeiner}}{{Schlafly} \&
  {Finkbeiner}}{2011}]{2011ApJ...737..103S}
{Schlafly} E.~F.,  {Finkbeiner} D.~P.,  2011, \mn@doi [\apj]
  {10.1088/0004-637X/737/2/103}, \href
  {https://ui.adsabs.harvard.edu/abs/2011ApJ...737..103S} {737, 103}

\bibitem[\protect\citeauthoryear{{Soria}, {Cropper}, {Pakull}, {Mushotzky}  \&
  {Wu}}{{Soria} et~al.}{2005}]{2005MNRAS.356...12S}
{Soria} R.,  {Cropper} M.,  {Pakull} M.,  {Mushotzky} R.,   {Wu} K.,  2005,
  \mn@doi [\mnras] {10.1111/j.1365-2966.2004.08438.x}, \href
  {https://ui.adsabs.harvard.edu/abs/2005MNRAS.356...12S} {356, 12}

\bibitem[\protect\citeauthoryear{{Soria}, {Kuntz}, {Long}, {Blair}, {Plucinsky}
   \& {Winkler}}{{Soria} et~al.}{2015}]{2015ApJ...799..140S}
{Soria} R.,  {Kuntz} K.~D.,  {Long} K.~S.,  {Blair} W.~P.,  {Plucinsky} P.~P.,
   {Winkler} P.~F.,  2015, \mn@doi [\apj] {10.1088/0004-637X/799/2/140}, \href
  {https://ui.adsabs.harvard.edu/abs/2015ApJ...799..140S} {799, 140}

\bibitem[\protect\citeauthoryear{{Stobbart}, {Roberts}  \&
  {Warwick}}{{Stobbart} et~al.}{2006}]{2006MNRAS.370...25S}
{Stobbart} A.~M.,  {Roberts} T.~P.,   {Warwick} R.~S.,  2006, \mn@doi [\mnras]
  {10.1111/j.1365-2966.2006.10494.x}, \href
  {https://ui.adsabs.harvard.edu/abs/2006MNRAS.370...25S} {370, 25}

\bibitem[\protect\citeauthoryear{{Stocke}, {Morris}, {Gioia}, {Maccacaro},
  {Schild}, {Wolter}, {Fleming}  \& {Henry}}{{Stocke}
  et~al.}{1991}]{1991ApJS...76..813S}
{Stocke} J.~T.,  {Morris} S.~L.,  {Gioia} I.~M.,  {Maccacaro} T.,  {Schild} R.,
   {Wolter} A.,  {Fleming} T.~A.,   {Henry} J.~P.,  1991, \mn@doi [\apjs]
  {10.1086/191582}, \href {http://adsabs.harvard.edu/abs/1991ApJS...76..813S}
  {76, 813}

\bibitem[\protect\citeauthoryear{{Sutton}, {Roberts}  \& {Middleton}}{{Sutton}
  et~al.}{2013}]{2013MNRAS.435.1758S}
{Sutton} A.~D.,  {Roberts} T.~P.,   {Middleton} M.~J.,  2013, \mn@doi [\mnras]
  {10.1093/mnras/stt1419}, \href
  {https://ui.adsabs.harvard.edu/abs/2013MNRAS.435.1758S} {435, 1758}

\bibitem[\protect\citeauthoryear{{Swartz}, {Soria}, {Tennant}  \&
  {Yukita}}{{Swartz} et~al.}{2011}]{2011ApJ...741...49S}
{Swartz} D.~A.,  {Soria} R.,  {Tennant} A.~F.,   {Yukita} M.,  2011, \mn@doi
  [\apj] {10.1088/0004-637X/741/1/49}, \href
  {http://adsabs.harvard.edu/abs/2011ApJ...741...49S} {741, 49}

\bibitem[\protect\citeauthoryear{{Tao}, {Feng}, {Gris{\'e}}  \& {Kaaret}}{{Tao}
  et~al.}{2011}]{2011ApJ...737...81T}
{Tao} L.,  {Feng} H.,  {Gris{\'e}} F.,   {Kaaret} P.,  2011, \mn@doi [\apj]
  {10.1088/0004-637X/737/2/81}, \href
  {http://adsabs.harvard.edu/abs/2011ApJ...737...81T} {737, 81}

\bibitem[\protect\citeauthoryear{{Thuan} \& {Martin}}{{Thuan} \&
  {Martin}}{1981}]{1981ApJ...247..823T}
{Thuan} T.~X.,  {Martin} G.~E.,  1981, \mn@doi [\apj] {10.1086/159094}, \href
  {https://ui.adsabs.harvard.edu/abs/1981ApJ...247..823T} {247, 823}

\bibitem[\protect\citeauthoryear{{Thuan}, {Bauer}  \& {Izotov}}{{Thuan}
  et~al.}{2014}]{2014MNRAS.441.1841T}
{Thuan} T.~X.,  {Bauer} F.~E.,   {Izotov} Y.~I.,  2014, \mn@doi [\mnras]
  {10.1093/mnras/stu716}, \href
  {https://ui.adsabs.harvard.edu/abs/2014MNRAS.441.1841T} {441, 1841}

\bibitem[\protect\citeauthoryear{{Tully} et~al.,}{{Tully}
  et~al.}{2013}]{2013AJ....146...86T}
{Tully} R.~B.,  et~al., 2013, \mn@doi [\aj] {10.1088/0004-6256/146/4/86}, \href
  {https://ui.adsabs.harvard.edu/abs/2013AJ....146...86T} {146, 86}

\bibitem[\protect\citeauthoryear{{Urquhart}, {Soria}, {Johnston}, {Pakull},
  {Motch}, {Schwope}, {Miller-Jones}  \& {Anderson}}{{Urquhart}
  et~al.}{2018}]{2018MNRAS.475.3561U}
{Urquhart} R.,  {Soria} R.,  {Johnston} H.~M.,  {Pakull} M.~W.,  {Motch} C.,
  {Schwope} A.,  {Miller-Jones} J.~C.~A.,   {Anderson} G.~E.,  2018, \mn@doi
  [\mnras] {10.1093/mnras/sty014}, \href
  {https://ui.adsabs.harvard.edu/abs/2018MNRAS.475.3561U} {475, 3561}

\bibitem[\protect\citeauthoryear{{Vierdayanti}, {Mineshige}, {Ebisawa}  \&
  {Kawaguchi}}{{Vierdayanti} et~al.}{2006}]{2006PASJ...58..915V}
{Vierdayanti} K.,  {Mineshige} S.,  {Ebisawa} K.,   {Kawaguchi} T.,  2006,
  \mn@doi [\pasj] {10.1093/pasj/58.5.915}, \href
  {https://ui.adsabs.harvard.edu/abs/2006PASJ...58..915V} {58, 915}

\bibitem[\protect\citeauthoryear{{Vinokurov}, {Fabrika}  \&
  {Atapin}}{{Vinokurov} et~al.}{2018}]{2018ApJ...854..176V}
{Vinokurov} A.,  {Fabrika} S.,   {Atapin} K.,  2018, \mn@doi [\apj]
  {10.3847/1538-4357/aaaa6c}, \href
  {http://adsabs.harvard.edu/abs/2018ApJ...854..176V} {854, 176}

\bibitem[\protect\citeauthoryear{{Vinokurov}, {Atapin}, {Solovyeva},
  {Sarkisyan}  \& {Oparin}}{{Vinokurov} et~al.}{2020}]{2020gbar.conf...95V}
{Vinokurov} A.,  {Atapin} K.,  {Solovyeva} Y.,  {Sarkisyan} A.,   {Oparin} D.,
  2020, in {Romanyuk} I.~I.,  {Yakunin} I.~A.,  {Valeev} A.~F.,   {Kudryavtsev}
  D.~O.,  eds, Ground-Based Astronomy in Russia. 21st Century. pp 95--99
  (\mn@eprint {arXiv} {2010.05866}),
  \mn@doi{10.26119/978-5-6045062-0-2_2020_95}

\bibitem[\protect\citeauthoryear{{Vogler} \& {Pietsch}}{{Vogler} \&
  {Pietsch}}{1997}]{1997A&A...319..459V}
{Vogler} A.,  {Pietsch} W.,  1997, \aap, \href
  {https://ui.adsabs.harvard.edu/abs/1997A&A...319..459V} {319, 459}

\bibitem[\protect\citeauthoryear{{Walton} et~al.,}{{Walton}
  et~al.}{2018a}]{2018ApJ...856..128W}
{Walton} D.~J.,  et~al., 2018a, \mn@doi [\apj] {10.3847/1538-4357/aab610},
  \href {https://ui.adsabs.harvard.edu/abs/2018ApJ...856..128W} {856, 128}

\bibitem[\protect\citeauthoryear{{Walton} et~al.,}{{Walton}
  et~al.}{2018b}]{2018ApJ...857L...3W}
{Walton} D.~J.,  et~al., 2018b, \mn@doi [\apjl] {10.3847/2041-8213/aabadc},
  \href {https://ui.adsabs.harvard.edu/abs/2018ApJ...857L...3W} {857, L3}

\bibitem[\protect\citeauthoryear{{Whitmore} et~al.,}{{Whitmore}
  et~al.}{2020}]{2020ApJ...889..154W}
{Whitmore} B.~C.,  et~al., 2020, \mn@doi [\apj] {10.3847/1538-4357/ab59e5},
  \href {https://ui.adsabs.harvard.edu/abs/2020ApJ...889..154W} {889, 154}

\bibitem[\protect\citeauthoryear{{Wiktorowicz}, {Sobolewska}, {Lasota}  \&
  {Belczynski}}{{Wiktorowicz} et~al.}{2017}]{2017ApJ...846...17W}
{Wiktorowicz} G.,  {Sobolewska} M.,  {Lasota} J.-P.,   {Belczynski} K.,  2017,
  \mn@doi [\apj] {10.3847/1538-4357/aa821d}, \href
  {https://ui.adsabs.harvard.edu/abs/2017ApJ...846...17W} {846, 17}

\bibitem[\protect\citeauthoryear{{Winter}, {Mushotzky}  \& {Reynolds}}{{Winter}
  et~al.}{2006}]{2006ApJ...649..730W}
{Winter} L.~M.,  {Mushotzky} R.~F.,   {Reynolds} C.~S.,  2006, \mn@doi [\apj]
  {10.1086/506579}, \href
  {https://ui.adsabs.harvard.edu/abs/2006ApJ...649..730W} {649, 730}

\bibitem[\protect\citeauthoryear{{Yang}, {Feng}  \& {Kaaret}}{{Yang}
  et~al.}{2011}]{2011ApJ.733.118Y}
{Yang} L.,  {Feng} H.,   {Kaaret} P.,  2011, \mn@doi [\apj]
  {10.1088/0004-637X/733/2/118}, \href
  {http://adsabs.harvard.edu/abs/2011ApJ...733..118Y} {733, 118}

\makeatother
\end{thebibliography}
%%%%%%%%%%%%%%%%%%%%%%%%%%%%%%%%%%%%%%%%%%%%%%%%%%

\begin{table*}
	\centering
	\caption{The log of {\it HST/ACS} and WFC3 observations of NGC 4861 and NGC 4449}
	\begin{tabular}{ l c c c c l l }
		\hline\hline
	 Source   & Filter & ObsID  & Date  & Exp.(ks)  \\
		\hline
 NGC 4861 &  WFC3/F487N & ibse02010 & 2012-10-29  & 2.68 \\
	&  WFC3/F502N & ibse02040 & 2012-10-29  & 1.20 \\
	&  WFC3/F547M &  ibse02050  & 2012-10-29  & 1.39 \\
	&  WFC3/F658N &  ibse02030  & 2012-10-29  & 1.18 \\
	&  WFC3/F673N &  ibse02020  & 2012-10-29  & 2.79 \\
	 &  WFC3/F814W & ibse02060 & 2012-10-29 & 1.04 \\
	 &  ACS/F606W & jc9v72010 & 2014-08-13 & 1.00  \\
	 &  ACS/F814W & jc9v72020  &  2014-08-13  & 1.00  \\ 		
 
		\hline
 NGC 4449  &  ACS/F435W & j9cd01040  & 2005-11-10 & 3.66  \\
    &  ACS/F555W & j9cd01020  &  2005-11-10  & 2.46  \\
	&  ACS/F814W & j9cd01030 & 2005-11-10 & 2.06 \\  
%	& & F658N & j9cd03020 & 2005-11-17  &  1.54 & \\ 
%   & & F814W & j9cd03010  & 2005-11-17 & 0.54 & \\ 
    &  ACS/F502N & j9dr01020  & 2005-11-18 &  1.28 &   \\
    &  ACS/F550M & j9dr01010 & 2005-11-18 & 1.20   &  \\ 
% & & F660N & j9dr01030  & 2005-11-18 & 1.86  &  \\ 
%  &  WFC3/F275W & icdm56050  & 2014-07-09 &  2.48 &   \\
%	&  WFC3/F336W & icdm56040 & 2014-07-09 & 2.36   &  \\
 \hline\hline
		\label{T:tab1}
	\end{tabular} 	
\end{table*}

\begin{table*}
	\centering
	\caption{The log of X-ray observations of NGC 4861 and NGC 4449}
	\begin{tabular}{ l c c c c  l l }
		\hline\hline
	 Source  & İnstruments  & ObsID  & Date  & Exp.(ks)  \\
		\hline
	  NGC 4861& {\it Chandra} &   12473 (C1)  & 2012-01-03 & 19.78  \\
 & &   19497 (C2)  & 2018-03-07  & 24.52   \\
		& &   20992 (C3)& 2018-03-11 & 58.46 \\   
		& &   20993 (C4)& 2018-03-16 & 27.69 \\ 
		& &   21036 (C5)& 2018-03-16 & 37.78 \\ 
	&{\it  XMM-Newton} &  0141150101 (XM1) & 2003-06-14 & 28.50  \\
		&  &  0141150401  (XM2)  & 2003-07-10  & 14.52   \\
		&  &  0141150501 (XM3) & 2003-12-03 & 21.85 \\
		\hline
	NGC 4449 & {\it Chandra}  &  2031 (C6)  & 2001-02-04 & 26.60  \\
		&  &  10125 (C7)  & 2009-03-04  &14.95   \\
		&  &   10875 (C8)& 2009-03-07 &59.39 \\
     &  {\it  XMM-Newton} &  0112521601 (XM4) & 2002-05-25 & 23.97  \\
		&  &  0112521701 (XM5)  & 2002-06-02  & 15.87   \\
		&  &  0824450801 (XM6) & 2018-05-11 & 87.00 \\
		&	{\it Swift}  & 00035873001 & 2007-03-27  & 4.71  \\	
	 &  & 00082261001 & 2013-11-19 & 2.61 \\
	 &  &  00082261003 & 2013-11-29 & 3.62 \\
	 &  & 00082261004 & 2013-12-01 & 1.05 \\	
	&  &   00082261006 & 2013-12-04 & 1.26 \\	
	&  &   00082261007 & 2013-12-06 & 4.96\\	
	&  &   00082261009 & 2013-12-10 & 1.23\\	
	&  &  00084419004  & 2016-09-30 & 0.79\\	
	&  &  00084419005  & 2017-10-01 & 0.71\\	
    
 \hline\hline
		\label{T:tab2}
	\end{tabular} 	
\end{table*}

\begin{table*}
\centering
\caption{Coordinates and their offsets of the X-Ray/Optical reference sources and position uncertainties of ULXs in NGC 4861.}
\begin{tabular}{ccccccccccccll}
\hline\hline
\multicolumn{6}{c}{{\it Chandra} ACIS X-ray sources (ObsID 20992) identified in {\it GAIA} sources }\\
\hline
&{\it Chandra} R.A.  & {\it Chandra} Dec.  & {\it GAIA} R.A.  & {\it GAIA} Dec.   & Offset \\
&(hh:mm:ss)  & ($\degr$:$\arcmin$:$\arcsec$) &  (hh:mm:ss)   & ($\degr$:$\arcmin$:$\arcsec$) & ($^{\prime\prime}$)\\
\hline
Ref1 & 12:59:04.068 & +34:47:02.63 & 12:59:03.934 & +34:47:02.40 & 1.67 \\
Ref2 & 12:58:59.364 & +34:49:42.72 & 12:58:59.232 & +34:49:42.45 & 1.64 \\
Ref3 & 12:59:04.061 & +34:53:53.61 & 12:59:03.914 & +34:53:53.51 & 1.84 \\
\hline
&{\it GAIA} R.A.& {\it GAIA} Dec.&{\it HST} R.A.&{\it HST} Dec.& Offset  \\
&(hh:mm:ss) &  ($\degr$:$\arcmin$:$\arcsec$) &  (hh:mm:ss)  &  ($\degr$:$\arcmin$:$\arcsec$) & ($^{\prime\prime}$) \\
\hline
Ref1 & 12:59:06.883 & +34:51:57.90 & 12:59:06.895 & +34:51:57.87 & 0.15 \\
Ref2 & 12:59:03.019 & +34:52:50.67 & 12:59:03.031 & +34:52:50.64 & 0.15 \\
Ref3 & 12:59:01.685 & +34:51:23.22 & 12:59:01.697 & +34:51:23.16 & 0.16 \\
Ref4 & 12:59:06.883 & +34:51:57.90 & 12:59:06.895 & +34:51:57.86 & 0.15 \\
Ref5 & 12:59:00.350 & +34:50:48.31 & 12:59:00.362 & +34:50:48.27 & 0.15 \\
\hline
\multicolumn{6}{c}{{\it Chandra} coordinates of the ULXs} {Corrected ULXs coordinates on {\it HST} image}\\
\hline
&R.A.  &  Dec. & R.A.  &  Dec.  & Position uncertainty\\
&(hh:mm:ss) &  ($\degr$:$\arcmin$:$\arcsec$) &(hh:mm:ss) &  ($\degr$:$\arcmin$:$\arcsec$) & ($\arcsec$)  \\
\hline
X1 & 12:59:01.949  & +34:51:14.79 & 12:59:01.822 & +34:51:14.54 & 0.24 \\
X2 & 12:59:00.965  & +34:50:47.90 & 12:59:00.839 & +34:50:47.66 & 0.24  \\
\hline
\end{tabular}
\label{T:tab3}
\end{table*}

\begin{table*}
\centering
\caption{Coordinates and their offsets of the X-Ray/Optical reference sources and position uncertainties of X-7 in NGC 4449.}
\begin{tabular}{ccccccccccccll}
\hline\hline
\multicolumn{6}{c}{{\it Chandra} ACIS X-ray sources (ObsID 20992) identified in {\it GAIA} sources }\\
\hline
&{\it Chandra} R.A.  & {\it Chandra} Dec.  & {\it HST} R.A.  & {\it HST} Dec.   & Offset \\
&(hh:mm:ss)  & ($\degr$:$\arcmin$:$\arcsec$) &  (hh:mm:ss)   & ($\degr$:$\arcmin$:$\arcsec$) & ($^{\prime\prime}$)\\
\hline
Ref1 & 12:28:11.990 & +44:06:41.65 & 12:28:12.062 & +44:06:42.71 & 1.31 \\
Ref2 & 12:28:15.665 & +44:05:37.05 & 12:28:15.728 & +44:05:38.17 & 1.31 \\
Ref3 & 12:28:07.353 & +44:04:54.08 & 12:28:07.429 & +44:04:55.08 & 1.29 \\
Ref4 & 12:28:01.369 & +44:05:30.33 & 12:28:01.439 & +44:05:31.35 & 1.27 \\
Ref5 & 12:28:10.960 & +44:06:48.66 & 12:28:11.030 & +44:06:49.70 & 1.28 \\
\hline
\multicolumn{6}{c}{{\it Chandra} coordinates of X7} {Corrected coordinates of X7 on {\it HST} image}\\
\hline
&R.A.  &  Dec. & R.A.  &  Dec.  & Position uncertainty\\
&(hh:mm:ss) &  ($\degr$:$\arcmin$:$\arcsec$) &(hh:mm:ss) &  ($\degr$:$\arcmin$:$\arcsec$) & ($\arcsec$)  \\
\hline
X7 & 12:28:17.831  & +44:06:34.10 & 12:28:17.901 & +44:06:35.15 & 0.20 \\
\hline
\end{tabular}
\label{T:tab3b}
\end{table*}

 \begin{table*}
	\centering
	\caption{The dereddened magnitude and color values of an optical candidate of X2 in NGC 4861.}
	\begin{tabular}{c c c c c c l l }
		\hline\hline
 & Filter & Vega Mag  &F$_x$ \\
 &     &  & (10$^{-15}$ erg cm$^{-2}$s$^{-1}$)   \\
		\hline
& WFC3/F487N &$23.07\pm0.19$  & 8.18  \\
 & WFC3/F502N & $23.73\pm0.48$  & 6.10 \\
 & WFC3/F547M & $23.86\pm0.08$ & 4.34 \\
 & WFC3/F658N &$22.24\pm0.59$  & 10.7$^{*}$      \\
 & WFC3/F673N & $24.03\pm0.27$ & 2.49 \\
 & WFC3/F814W  & $23.99\pm0.12$ & 1.74 \\
& ACS/F606W(V)  & $23.80\pm0.04$ & 4.16  \\  
 & ACS/F814W(I) & $24.14\pm0.08$ & 1.70   \\
  &$(V-I)_{0}$ & -0.34 &  & \\
 &$M_{v}$  & -6.2 & & \\
		\hline	\hline
		\end{tabular}
		\label{T:tab4}
		\\ \footnotesize{$^{*}$ see section \ref{sec:res} for an explanation of this flux value.}
\end{table*}

\begin{table*}
	\centering
	\caption{The dereddened magnitude and color values of optical candidates of X7 in NGC 4449.}
	\begin{tabular}{c c c c c c c c c c  l l }
		\hline\hline
& Source & Filter & Vega Mag & Johnson Mag \\
&   &      &  & &  \\
		\hline
& A1 & ACS/F435W (B) & 22.89$\pm$0.02 &  22.94$\pm$0.02 & \\
&  & ACS/F555W (V) & 22.89$\pm$0.02  & 22.86$\pm$0.02 &  \\
&  & ACS/F814W (I)  & 23.29$\pm$0.03  & 23.28$\pm$0.03  &  \\
%&  & WFC3/F336W  &22.11$\pm$0.05&  -   \\
%&  & WFC3/F275W  & 20.99$\pm$0.06&  -   \\ 
& & $(B-V)_{0}$ & 0.0 & 0.08 & & & \\
& & $M_{v}$ & -5.00 & -5.00  & & & \\

& A2 & ACS/F435W & 23.29$\pm$0.03  & 23.31$\pm$0.03   & \\
&  & ACS/F555W & 22.91$\pm$0.03  &  22.87$\pm$0.03  &  \\
&  & ACS/F814W  & 23.28$\pm$0.04   & 23.27$\pm$0.04  &  \\
%&  & WFC3/F336W  & 21.98$\pm$0.05  &  -  & \\
%&  & WFC3/F275W  & 21.71$\pm$0.07  &  -  &  \\ 
& & $(B-V)_{0}$ & 0.38  & 0.44  & &  &\\
& & $M_{v}$ & -5.00 &-5.00 & &  &\\

& A3 & ACS/F435W & 23.66$\pm$0.03  & 23.71$\pm$0.03  &  \\
&  & ACS/F555W & 23.74$\pm$0.04  & 23.72$\pm$0.04 &  \\
&  & ACS/F814W  & 24.01$\pm$0.06 & 24.00$\pm$0.04  & \\
%&  & WFC3/F336W  & 22.21$\pm$0.05 &  -  & \\
%&  & WFC3/F275W  & 21.68$\pm$0.07  & - &  \\
& & $(B-V)_{0}$ & -0.08  & -0.01 & &  &\\
& & $M_{v}$ &  -4.10 & -4.10 & & & \\
		\hline	\hline
		\end{tabular}
		\label{T:tab5}	
\end{table*} 

\begin{table*}
	\centering
	\begin{minipage}[b]{0.9\linewidth}
	\caption{X-ray spectral parameters of NGC 4861 X1 and X2 in the {\it XMM-Newton} and {\it Chandra} observations }
	\begin{tabular}{c c c c c c c r r r l }
	\hline\hline
 No.  &  model &   $\Gamma$ &  kT  & p & N$_{PL}$ \footnote{Normalization parameters of {\scshape pl} model in units of photon keV$^{-1}$ cm$^{-2}$ s$^{-1}$ at 1 keV}   & N$_{Disk}$ \footnote{Normalization parameters of {\scshape dıskpbb} and {\scshape diskbb} models; N$_{Disk}$=[(r$_{in}$ km$^{-1}$)/(D/10 kpc)]$^{2}$ $\times cosi$.} & F (10$^{-13}$) & L$_{x}$\footnote{Luminosity values were calculated at 0.3$-$10 keV energy range. Errors indicate the 90$\%$ confidence interval for each parameter.} (10$^{39}$) & C/dof \footnote{C-statistic of the fit and the number of degrees of freedom.} \\
	&       & &   keV &       &  $10^{-4}$   &  $10^{-4}$  & erg cm$^{-2}$s$^{-1}$  & erg s$^{-1}$ &\\
\hline
	  &    &    &    &  X1    \\
\hline
	
C1 &{\scshape pl}  & $1.98_{-0.10}^{+0.11}$ & $-$  & & $1.22_{-0.06}^{+0.06}$ & & $6.91_{-0.02}^{+0.02}$ & $8.17_{-0.02}^{+0.02}$ & 80.35/67 \\
   & {\scshape dıskpbb} & $-$ & $2.15_{-0.03}^{+0.03}$  & $0.5_{-0.04}^{+0.04}$ & & $2.89_{-0.15}^{+0.15}$  & $6.31_{-0.06}^{+0.06}$ & $7.47_{-0.06}^{+0.07}$ & 78.60/66\\
   & {\scshape pl+dıskpbb} &$0.40_{-0.14}^{+0.15}$ & $0.81_{-0.01}^{+0.01}$  & $0.5_{-0.009}^{+0.01}$ &$0.06_{-0.01}^{+0.01}$  &$207_{-13.39}^{+13.39}$  & $6.88_{-0.02}^{+0.02}$ & $8.15_{-0.11}^{+0.29}$ & 74.81/64\\ 
   
C2 & {\scshape pl} &  $2.48_{-0.14}^{+0.15}$ & $-$   & & $0.48_{-0.06}^{+0.06}$ & & $2.35_{-0.05}^{+0.05}$ & $2.78_{-0.03}^{+0.03}$ & 12.46/14 \\
   & {\scshape dıskpbb} & $-$ & $1.03_{-0.02}^{+0.02}$  & $0.5_{*}^{+0.08}$ & & $14.4_{-1.24}^{+1.32}$  & $1.83_{-0.04}^{+0.03}$ & $2.17_{-0.06}^{+0.10}$ & 19.58/13 \\
   & {\scshape pl+dıskpbb} & $1.95_{-0.14}^{+0.15}$ & $0.21_{-0.01}^{+0.009}$  & $0.9_{-0.16}^{+0.93}$ &$0.28_{-0.03}^{+0.03}$ & * & $3.39_{-0.03}^{+0.03}$ & $4.02_{-0.04}^{+0.05}$ & 9.50/11\\

C3 & {\scshape pl} & $2.32_{-0.08}^{+0.08}$ & $-$  & & $0.46_{-0.02}^{+0.02}$ & &$2.28_{-0.02}^{+0.02}$ & $2.70_{-0.02}^{+0.03}$ & 83.17/71 \\
  	& {\scshape dıskpbb}  & $-$ & $1.30_{-0.01}^{+0.01}$  & $0.5_{*}^{+0.03}$ & & $5.71_{-0.24}^{+0.24}$  & $1.91_{-0.02}^{+0.02}$ &$2.26_{-0.03}^{+0.03}$ & 79.76/70 \\	
    & {\scshape pl+dıskpbb} & $2.50_{-0.09}^{+0.09}$ & $0.90_{-0.04}^{+0.04}$  & $0.8_{-0.05}^{+0.07}$ & $0.35_{-0.02}^{+0.02}$& $40.0_{-7.30}^{+7.30}$ & $2.02_{-0.02}^{+0.03}$ & $2.39_{-0.02}^{+0.02}$ & 79.12/68\\

C4  & {\scshape pl} & $2.39_{-0.12}^{+0.13}$ & $-$& & $0.48_{-0.03}^{+0.03}$ & & $2.35_{-0.04}^{+0.03}$ & $2.78_{-0.04}^{+0.03}$ & 30.66/34 \\
 	& {\scshape dıskpbb} & $-$ & $1.39_{-0.02}^{+0.02}$  & $0.5_{*}^{+0.05}$ & &$4.23_{-0.33}^{+0.33}$ & $1.90_{-0.05}^{+0.03}$  & $2.25_{-0.05}^{+0.04}$ & 32.30/33\\
 	& {\scshape pl+dıskpbb} & $2.39_{-0.12}^{+0.14}$ & $0.10_{-0.1}^{+0.03}$  & $0.5_{*}^{+0.05}$ &$0.48_{-0.03}^{+0.03}$ & $0.0001_{*}^{*}$ & $3.88_{-0.03}^{+0.03}$ & $4.59_{-0.04}^{+0.05}$ & 30.67/31\\

C5  & {\scshape pl} & $2.35_{-0.02}^{+0.02}$ & -  & & $0.46_{-0.03}^{+0.03}$ & & $2.26_{-0.03}^{+0.03}$ & $2.53_{-0.03}^{+0.03}$ & 100.23/84 \\
	& {\scshape dıskpbb} &$-$ & $1.64_{-0.03}^{+0.03}$ & $0.5_{*}^{+0.04}$ & & $2.15_{-0.14}^{+0.14}$ & $2.14_{-0.03}^{+0.03}$ & $2.53_{-0.05}^{+0.04}$ & 108.91/83  \\
    & {\scshape pl+dıskpbb} & $1.88_{-0.04}^{+0.04}$ & $0.26_{-0.01}^{+0.02}$  & $0.9_{-0.01}^{+0.09}$ &$0.26_{-0.02}^{+0.02}$ & {*} &  $2.17_{-0.03}^{+0.04}$ & $2.57_{-0.04}^{+0.03}$& 95.10/81 \\ 
    
	XM1 & {\scshape pl}  & $1.89_{-0.04}^{+0.05}$ &$-$ &  & $0.51_{-0.02}^{+0.02}$ & & $2.78_{-0.01}^{+0.01}$ & $3.29_{-0.02}^{+0.02}$ & 335.18/229 \\	
    & {\scshape dıskpbb} &$-$ & $1.31_{-0.01}^{+0.01}$  &  $0.5_{-0.004}^{+0.004}$ & & $13.4_{-0.67}^{+0.67}$ & $2.29_{-0.01}^{+0.01}$ & $2.71_{-0.01}^{+0.02}$ & 308.44 /228\\
    & {\scshape pl+dıskpbb}  & $1.26_{-0.08}^{+0.08}$ & $0.40_{-0.007}^{+0.007}$  & $1_{*}^{+0.03}$ &$0.16_{-0.01}^{+0.01}$  & $*$  & $2.64_{-0.01}^{+0.01}$ & $3.12_{-0.02}^{+0.02}$ & 260.07/226\\

XM2 & {\scshape pl} & $1.89_{-0.11}^{+0.11}$ & $-$  & & $0.42_{-0.04}^{+0.04}$ & & $2.53_{-0.05}^{+0.05}$ & $2.99_{-0.05}^{+0.06}$  & 37.28/41\\  
    & {\scshape dıskpbb} & $-$ & $1.83_{-0.05}^{+0.05}$  & $0.5_{-0.01}^{+0.03}$ &  & $2.13_{-0.22}^{+0.21}$ & $2.36_{-0.05}^{+0.05}$ & $2.79_{-0.05}^{+0.06}$ & 36.70/40 \\
    & {\scshape pl+dıskpbb} & $1.90_{-0.12}^{+0.12}$ & $0.10_{*}^{+0.10}$  & $0.9_{*}^{+0.98}$ & $0.42_{-0.04}^{+0.04}$& $*$& $2.35_{-0.05}^{+0.05}$ & $2.79_{-0.06}^{+0.05}$ & 37.28/38\\

XM3 & {\scshape pl} &  $2.07_{-0.13}^{+0.13}$ & $-$&  &$0.29_{-0.03}^{+0.03}$ & & $1.61_{-0.05}^{+0.05}$ & $1.90_{-0.05}^{+0.06}$ & 95.69/96 \\  
   & {\scshape dıskpbb} &$-$ & $2.04_{-0.06}^{+0.06}$  & $0.5_{*}^{+0.05}$ & & $0.62_{-0.07}^{+0.07}$ & $1.77_{-0.03}^{+0.02}$ & $1.90_{-0.03}^{+0.03}$ &96.94/95\\
   & {\scshape pl+dıskpbb}  & $1.39_{-0.17}^{+0.20}$ & $0.27_{-0.01}^{+0.01}$  & $0.9_{-0.08}^{+0.08}$ &$0.13_{-0.02}^{+0.02}$ &$*$ & $1.74_{-0.05}^{+0.04}$ & $2.06_{-0.04}^{+0.05}$ & 88.37/93 \\ 

\hline
	&   &    &   & X2 \\
\hline
	
C1  & {\scshape pl}  & $2.11_{-0.23}^{+0.26}$ & $-$  & & $0.25_{-0.03}^{+0.03}$ & & $1.33_{-0.07}^{+0.06}$ & $1.58_{-0.07}^{+0.06}$ & 13.11/16   \\
    & {\scshape dıskbb} & $-$ & $0.81_{-0.03}^{+0.03}$   &  & & $95.4_{-14.40}^{+14.40}$ & $0.86_{-0.04}^{+0.05}$ & $1.01_{-0.03}^{+0.04}$& 12.32/16 \\
  & {\scshape pl+dıskbb} &  $6.46_{-4.74}^{+6.60}$ & $0.89_{-0.04}^{+0.03}$  & & $0.03_{-0.03}^{+0.04}$& $63.7_{-9.80}^{+9.90}$&  $3.36_{-0.06}^{+0.07}$ & $3.97_{-0.06}^{+0.06}$ & 11.04/14 \\   
	    
C2  & {\scshape pl}  & $1.84_{-0.15}^{+0.16}$ & $-$  & &$0.30_{-0.03}^{+0.03}$ & & $1.89_{-0.04}^{+0.04}$ & $2.24_{-0.05}^{+0.05}$  & 17.25/15\\
  	& {\scshape dıskbb} & $-$ & $0.87_{-0.02}^{+0.02}$   & & & $94.0_{-9.22}^{+9.22}$ & $1.16_{-0.03}^{+0.04}$ & $1.37_{-0.03}^{+0.04}$ & 12.60/15\\
    & {\scshape pl+dıskbb} &  $0.58_{-2.08}^{+0.51}$ & $0.89_{-0.02}^{+0.02}$  & & $0.01_{-0.01}^{+0.01}$& $101_{-9.45}^{+9.45}$&  $1.89_{-0.04}^{+0.04}$ & $2.24_{-0.05}^{+0.05}$ & 12.63/13\\ 
 
C3  & {\scshape pl}  & $2.10_{-0.08}^{+0.08}$ & $-$  & &$0.33_{-0.02}^{+0.02}$ & & $1.80_{-0.02}^{+0.02}$ & $2.13_{-0.02}^{+0.03}$ &155.50/101 \\
	& {\scshape dıskbb} &$-$  & $0.80_{-0.01}^{+0.01}$  &  & & $146_{-8.72}^{+8.72}$  & $1.19_{-0.02}^{+0.02}$ & $1.41_{-0.03}^{+0.02}$ & 106.50/101  \\
	& {\scshape pl+dıskbb}& $0.44_{-0.70}^{+6.20}$ & $0.77_{-0.01}^{+0.01}$  & & $0.001_{-0.001}^{+0.001}$ & $160_{-9.67}^{+9.67}$ &  $1.29_{-0.02}^{+0.02}$ & $1.53_{-0.03}^{+0.02}$ & 106.22/99\\ 
	
C4  & {\scshape pl} &$2.01_{-0.11}^{+0.12}$ & $-$ & &$0.32_{-0.04}^{+0.04}$ & & $1.81_{-0.03}^{+0.03}$ & $2.15_{-0.04}^{+0.03}$  & 22.68/19 \\
	& {\scshape dıskbb} &  $-$ & $0.95_{-0.02}^{+0.02}$ &  & & $73.6_{-6.48}^{+6.88}$ & $1.21_{-0.04}^{+0.03}$  & $1.44_{-0.05}^{+0.04}$ & 10.80/19\\
    & {\scshape pl+dıskbb} & $0.22_{-0.36}^{+0.69}$ & $0.83_{-0.02}^{+0.02}$  &  &$0.005_{-0.003}^{+0.003}$  &$116_{-10.8}^{+11.5}$  & $1.41_{-0.04}^{+0.03}$ & $1.67_{-0.05}^{+0.04}$ & 9.26/17\\

C5 & {\scshape pl} & $2.04_{-0.10}^{+0.10}$ & $-$ & & $0.33_{-0.02}^{+0.02}$  & & $1.83_{-0.03}^{+0.03}$ & $ 2.17_{-0.04}^{+0.04}$ & 129.41/75 \\
   	& {\scshape dıskbb} &$-$  & $0.86_{-0.01}^{+0.01}$  &  & & $112_{-8.32}^{+8.32}$ &$1.23_{-0.03}^{+0.03}$ & $1.46_{-0.04}^{+0.03}$  & 88.71/75 \\
    & {\scshape pl+dıskbb} & $9.49_{-9.49}^{+9.49}$ & $0.85_{-0.01}^{+0.01}$  & & {*}  & $114_{-8.41}^{+8.82}$ &  $1.22_{-0.03}^{+0.03}$ & $1.45_{-0.03}^{+0.04}$ & 88.71/73\\
   
XM1 & {\scshape pl} &  $1.96_{-0.08}^{+0.08}$ & $-$ & & $0.28_{-0.02}^{+0.02}$  &  & $1.61_{-0.03}^{+0.03}$ & $ 1.90_{-0.04}^{+0.03}$ & 124.87/83 \\
    & {\scshape dıskbb}  & $-$ & $0.63_{-0.01}^{+0.01}$   & & &$346_{-25.2}^{+25.2}$ & $1.09_{-0.03}^{+0.03}$ & $1.30_{-0.03}^{+0.04}$  &90.14/83\\   	
   	& {\scshape pl+dıskbb}  & $1.70_{-0.20}^{+0.24}$ & $0.56_{-0.01}^{+0.01}$ &  &$0.07_{-0.01}^{+0.01}$  & $395_{-31.5}^{+31.5}$ &  $1.29_{-0.03}^{+0.03}$ & $1.53_{-0.04}^{+0.03}$ & 79.56/81\\

XM2 & {\scshape pl} & $2.06_{-0.14}^{+0.15}$ & $-$ &  & $0.26_{-0.03}^{+0.03}$  &  & $1.61_{-0.03}^{+0.03}$ & $ 1.91_{-0.04}^{+0.03}$ &36.53/27 \\
   	& {\scshape dıskbb}  & $-$ & $0.61_{-0.02}^{+0.02}$  &  &  & $357_{-48.7}^{+48.7}$ & $1.00_{-0.05}^{+0.05}$ & $1.19_{-0.05}^{+0.04}$ & 41.48/27 \\   	
	& {\scshape pl+dıskbb} & $2.06_{-0.14}^{+0.15}$ & $6.32_{-0.06}^{+1.64}$ &  & $0.26_{-0.03}^{+0.03}$ & $0.14_{-0.14}^{+0.14}$  & $1.21_{-0.06}^{+0.06}$ & $1.44_{-0.06}^{+0.06}$ & 36.53/25\\  
	
	\hline\hline
	\end{tabular}
\\$^{*}$:The parameter was unconstrained\\
	\label{T:tab7}
	\end{minipage}
\end{table*}

\begin{table*}
	\centering
	\begin{minipage}[b]{0.9\linewidth}
	\caption{X-ray spectral parameters of NGC 4449 X7 in the {\it XMM-Newton} and {\it Chandra} observations }
	\begin{tabular}{c c c c c c c r r r l }
	\hline\hline
 No.  &  model & N$_{H}$ &  $\Gamma$ &  kT   & N$_{PL}$ \footnote{Normalization parameters of {\scshape pl} model in units of photon keV$^{-1}$ cm$^{-2}$ s$^{-1}$ at 1 keV}   & N$_{Disk}$ \footnote{Normalization parameters of {\scshape dıskpbb} and {\scshape diskbb} models; N$_{Disk}$=[(r$_{in}$ km$^{-1}$)/(D/10 kpc)]$^{2}$ $\times cosi$.} & F (10$^{-13}$) & L$_{x}$\footnote{Luminosity values were calculated at 0.3$-$10 keV energy range. Errors indicate the 90$\%$ confidence interval for each parameter.} (10$^{39}$) & C/dof \footnote{C-statistic of the fit and the number of degrees of freedom.}\\
	&      & $(10^{22})$cm$^{-2}$  &    keV & &    $10^{-4}$   &  $10^{-4}$  & erg cm$^{-2}$s$^{-1}$  & erg s$^{-1}$  &\\
\hline

C6 & {\scshape pl} & $0.59_{-0.03}^{+0.03}$ & $1.97_{-0.05}^{+0.05}$ & $-$  & $1.73_{-0.07}^{+0.07}$ &  & $9.84_{-0.01}^{+0.01}$ & $ 1.90_{-0.02}^{+0.03}$ & 53.76/56 \\
   & {\scshape dıskbb} & $0.29_{-0.03}^{+0.03}$ & $-$ & $1.50_{-0.01}^{+0.01}$  & & $58.0_{-2.53}^{+2.60}$ & $6.02_{-0.01}^{+0.01}$ & $1.16_{-0.01}^{+0.02}$ & 80.42/56 \\
   & {\scshape pl+dıskbb} & $0.75_{-0.03}^{+0.03}$ & $2.79_{-0.09}^{+0.10}$ & $4.86_{-0.20}^{+0.19}$  &  $2.43_{-0.14}^{+0.14}$ & $0.50_{-0.08}^{+0.08}$ & $13.4_{-0.01}^{+0.01}$ & $2.59_{-0.02}^{+0.03}$ & 51.37/54\\  
   
C7 & {\scshape pl} & $0.69_{-0.08}^{+0.08}$ & $2.62_{-0.13}^{+0.14}$ & $-$ & $1.06_{-0.10}^{+0.10}$ & & $5.10_{-0.04}^{+0.04}$ & $ 0.98_{-0.05}^{+0.04}$ & 24.89/21 \\
   & {\scshape dıskbb} & $0.27_{-0.07}^{+0.08}$ & $-$ & $0.98_{-0.02}^{+0.02}$   & & $97.0_{-9.53}^{+9.53}$ & $1.84_{-0.04}^{+0.04}$ & $0.35_{-0.06}^{+0.04}$ & 23.57/21 \\  
   & {\scshape pl+dıskbb} & $0.69_{-0.08}^{+0.08}$ & $2.62_{-0.13}^{+0.15}$ & $9.92_{*}^{*}$ &  $1.06_{-0.10}^{+0.10}$ & $0.66_{-0.66}^{+0.66}$ & $2.32_{-0.04}^{+0.04}$ & $0.45_{-0.04}^{+0.05}$ &  24.89/19\\   
   	
C8  & {\scshape pl} & $0.80_{-0.03}^{+0.03}$ & $2.89_{-0.06}^{+0.06}$ & $-$ & $1.78_{-0.07}^{+0.07}$ &  & $8.97_{-0.01}^{+0.01}$ & $ 1.73_{-0.02}^{+0.02}$ & 65.10/54 \\
    & {\scshape dıskbb} & $0.36_{-0.03}^{+0.03}$ & $-$ & $0.87_{-0.09}^{+0.09}$   & & $213_{-9.38}^{+9.66}$ &$2.60_{-0.01}^{+0.01}$ & $0.50_{-0.03}^{+0.02}$ & 67.97/54 \\ 
   & {\scshape pl+dıskbb} & $0.67_{-0.03}^{+0.03}$ & $2.85_{-0.10}^{+0.10}$ & $0.80_{-0.02}^{+0.02}$  & $0.99_{-0.06}^{+0.06}$ & $134_{-16.0}^{+16.0}$ & $2.62_{-0.01}^{+0.01}$ & $0.50_{-0.02}^{+0.02}$ & 61.44/52\\    
   	
XM4   & {\scshape pl} & $0.52_{-0.06}^{+0.06}$& $2.04_{-0.11}^{+0.12}$ & $-$  &$1.70_{-0.22}^{+0.22}$  & & $9.32_{-0.06}^{+0.05}$ & $1.80_{-0.05}^{+0.07}$ & 61.75/53\\
    & {\scshape dıskbb} & ${0.23_{-0.04}^{+0.05}}$ & $-$ & ${1.36_{-0.02}^{+0.02}}$  &  &$78.7_{-5.05}^{+5.05}$ & $5.34_{-0.06}^{+0.05}$ & $1.03_{-0.06}^{+0.06}$ & 68.84/53 \\	
    & {\scshape pl+dıskbb} & $0.53_{-0.06}^{+0.06}$ & $2.00_{-0.11}^{+0.12}$ & $0.40_{-0.12}^{+0.08}$  & $1.60_{-0.20}^{+0.20}$  & $571_{-7.35}^{+7.35}$ & $9.32_{-0.09}^{+0.10}$  & $1.80_{-0.10}^{+0.10}$ & 61.73/51\\
    
XM5   & {\scshape pl} & $0.61_{-0.03}^{+0.04}$& $2.31_{-0.06}^{+0.06}$ & $-$ &$2.66_{-0.15}^{+0.15}$ & & $12.4_{-0.04}^{+0.05}$ & $2.40_{-0.05}^{+0.05}$  & 85.28/73\\
	& {\scshape dıskbb} & ${0.26_{-0.03}^{+0.03}}$ & $-$ & ${1.19_{-0.01}^{+0.02}}$  & & $150_{-8.64}^{+8.64}$ & $6.02_{-0.01}^{+0.01}$ & $1.16_{-0.04}^{+0.03}$ & 104.83/73\\
	& {\scshape pl+dıskbb} & $0.85_{-0.03}^{+0.04}$ & $3.87_{-0.18}^{+0.20}$ & $1.69_{-0.04}^{+0.04}$  &$4.51_{-0.30}^{+0.30}$  & $26.9_{-2.79}^{+2.79}$ & $20.0_{-0.35}^{+0.35}$  & $3.87_{-0.36}^{+0.35}$ & 83.70/71\\

XM6 & {\scshape pl} & $0.34_{-0.01}^{+0.01}$& $1.86_{-0.02}^{+0.02}$ & $-$  &$1.10_{-0.02}^{+0.02}$  & & $6.60_{-0.01}^{+0.01}$ & $1.27_{-0.01}^{+0.01}$ & 422.34/406 \\
	& {\scshape dıskbb} & ${0.13_{-0.008}^{+0.009}}$ & $-$ & ${1.57_{-0.01}^{+0.01}}$  & & $36.6_{-0.84}^{+0.84}$ & $4.58_{-0.01}^{+0.01}$ & $0.86_{-0.01}^{+0.03}$ & 559.98/406\\
	& {\scshape pl+dıskbb} & ${0.32_{-0.009}^{+0.008}}$ & $1.79_{-0.02}^{+0.02}$ & $0.62_{-0.05}^{+0.05}$  &$1.00_{-0.02}^{+0.02}$  & $0.96_{-0.02}^{+0.02}$ & $4.49_{-0.01}^{+0.01}$  & $0.86_{-0.01}^{+0.01}$ & 422.04/404\\

	\hline\hline
	\end{tabular}
\\$^{*}$:The parameter was unconstrained\\
Note: Only the X-ray absorption value $N_{H}=1.1\times10^{20}cm^{-2}$ was used and this value was kept fixed.
	\label{T:tab8}
	\end{minipage}
\end{table*}

%%%%%%%%%%%%%%%%%%%%%%%%%%%%%%%%%%%%%%%%%%%%%%%%%%

% Don't change these lines
\bsp	% typesetting comment
\label{lastpage}
\end{document}